\documentclass[draftcls,onecolumn,12pt]{IEEEtran}

\usepackage{cite}

\ifCLASSINFOpdf
\else
   \usepackage[dvips]{graphicx}
\fi
\usepackage[dvips]{epsfig,graphics,graphicx}
\usepackage{psfrag}
\usepackage{subfigure}

\usepackage[cmex10]{amsmath}
\usepackage{amsthm,amsfonts,amssymb}
\usepackage{xspace}
\usepackage{setspace}

\usepackage{mdwmath}
\usepackage{mdwtab}
\usepackage{commands}

\newcommand{\sqr}[1]{\sqrt{#1}}
\newcommand{\sqrr}[1]{{\sqrt[3]{#1}}}
\newcommand{\s}{S_X(e^{j\omega})}
\newcommand{\sx}{S_X(e^{j\omega})}
\newcommand{\sxx}[1]{S_X^{#1}(e^{j\omega})}

\newcommand{\Tp}{\Theta_+(\ej)}
\newcommand{\Tm}{\Theta_-(\ej)}
\newcommand{\Ds}{D_S(\ej)}
\newcommand{\Dc}{D_C(\ej)}

\newcommand{\tp}{\Theta_+}
\newcommand{\tm}{\Theta_-}
\renewcommand{\l}{\mathcal{L}_{\lambda_1,\lambda_2}(\eta,\tp,\tm)}
\newcommand{\tpk}{\tp(k)}
\newcommand{\tmk}{\tm(k)}
\newcommand{\lk} {\tilde{\mathcal{L}}_{\lambda_1,\lambda_2,k}(\eta(k),\tpk,\tmk)}

\newcommand{\rem}[1]{}

\usepackage{times}

\begin{document}
%
\title{Colored-Gaussian Multiple Descriptions: \\ Spectral and Time-Domain Forms}

%
%
%

\author{Jan {\O}stergaard,~\IEEEmembership{Senior Member,~IEEE,}
 Yuval Kochman,~\IEEEmembership{Member,~IEEE,}
         and~Ram~Zamir,~\IEEEmembership{Fellow,~IEEE}
 \thanks{This work was presented in part at the IEEE Data Compression Conference, Snowbird, Utah, 2008.}
\thanks{J.~{\O}stergaard (jo@es.aau.dk) is with the Department of Electronic Systems, Aalborg University, Aalborg, Denmark.
The work of J. {\O}stergaard was partially supported by VILLUM FONDEN Young Investigator Programme, Project No.\ 10095.}%
 \thanks{Y.~Kochman (yuvalko@mit.edu) is with the School of Computer Science and Engineering, The Hebrew University of Jerusalem, Jerusalem, Israel.
  The work of~Y.~Kochman was partially supported by the Israel Science Foundation, grant no.\ 956/12.}%
 \thanks{R.\ Zamir (zamir@eng.tau.ac.il) is with the Department of Electrical Engineering-Systems, Tel Aviv University, Tel Aviv, Israel.
The work of R.~Zamir was partially supported by ISF grant 870/11.}
\thanks{Copyright (c) 2014 IEEE. Personal use of this material is permitted.  However, permission to use this material for any other purposes must be obtained from the IEEE by sending a request to pubs-permissions@ieee.org.}}

%
%

\markboth{Accepted for publication in the IEEE Transactions on Information Theory, December 2015}{}
%



\maketitle

\begin{abstract}
It is well known that Shannon's rate-distortion function (RDF) in the colored quadratic Gaussian (QG) case can be parametrized via a single Lagrangian variable (the ``water level'' in the reverse water filling solution).
In this work, we show that the symmetric colored QG
multiple-description (MD) RDF in the case of two descriptions can be
parametrized  in the spectral domain
via two Lagrangian variables, which control the trade-off between the side distortion, the central distortion, and the coding rate.
This spectral-domain analysis is complemented by a time-domain scheme-design approach: 
we show that the symmetric colored QG MD RDF can be achieved by combining
ideas of delta-sigma modulation and differential pulse-code modulation. Specifically,
two source prediction loops, one for each description, are embedded within a common
noise shaping loop, whose parameters are explicitly found from the
spectral-domain characterization.

\end{abstract}



%
\IEEEpeerreviewmaketitle

\section{Introduction}
\IEEEPARstart{T}{he}
traditional multiple-description (MD) problem
\cite{elgamal:1982,ozarow:1980} considers a source, which is encoded
into two descriptions that are transmitted over separate
channels. Either one of the channels may break down and thereby cause
a description loss at the decoder. A description is either received
error-free at the decoder or not received at all. While the encoder does not
know which of the channels that are working, it is assumed that the
decoder can identify the received descriptions.
The problem is then to design the two descriptions so that they individually represent the source to within some desired distortion level and yet are able to refine each other. Thus, combining the descriptions improves upon their individual performances.
In the \emph{symmetric} situation, the two descriptions are balanced, i.e., they are encoded at the same coding rate, and lead to the same distortion when used separately at the decoder.




The achievable rate-distortion region for the MD problem is only
completely known for very few cases.
El-Gamal and Cover~\cite{elgamal:1982} presented an achievable rate region for two
descriptions and memoryless sources. Ozarow~\cite{ozarow:1980}
showed that in the white quadratic Gaussian (QG) case, i.e., for white
Gaussian sources and mean-squared error (MSE) distortion, the El-Gamal and Cover
region is tight.
For the case of general (time-correlated) stationary Gaussian sources and MSE distortion, Dragotti et al. \cite{Dragotti:2002} characterized the achievable rate region in the high-resolution limit. The region at general resolution was recently characterized by
Chen et al.~\cite{chen:2009}.
In particular, it was shown in~\cite{chen:2009} that the achievable rate region forms a closed and convex set and that the minimal description rates can be found by extremizing over all distortion spectra satisfying the individual side and central distortion constraints. 

The results in~\cite{chen:2009} do not specify an explicit solution to the optimal
distortion spectra for the colored QG MD problem. Nevertheless, they provide some
intuition towards a spectral domain characterization.  Specifically, they
show that the optimal rates for stationary Gaussian sources
can be expressed as the sum of rates of parallel channels, each one
representing a frequency band. Each of the channels must be tuned to a
minimum Ozarow MD rate for some frequency dependent distortion
level. In some sense, this can be seen as a reverse ``water-filling''
approach, where instead of having a \emph{flat} water level as in the
conventional single-description (SD) case, the water level may be
frequency dependent. The authors of~\cite{chen:2009} also pointed out that obtaining an
explicit spectral domain characterization from their results is technically
non-trivial, since their result is given in the form of a minimization problem, which does not seem to have an explicit solution.
Instead it was argued that the optimal rates can be found through numerical optimization by approximating the source spectral density by piece-wise constant functions. However, in general, for arbitrarily shaped sources, this becomes an infinite-dimensional optimization problem.

In the first part of this paper,
we present a parametrization of the symmetric colored QG MD rate-distortion function (RDF). While Shannon's RDF in the SD case can be parametrized by a single Lagrangian variable~\cite{kolmogorov:1956} (usually referred to as a ``water level''), the symmetric colored
QG MD RDF can be parametrized via \emph{two} Lagrangian variables.\footnote{In our case, however, the two parameters cannot generally be interpreted as ``water levels''.}
To establish this result, we use two key ideas.
We propose a new ``differential'' representation for the MD test channel
(see e.g., Fig.~\ref{DPCM_DSM_2}), and show that the mutual information rate
across this channel coincides with the QG MD RDF.
%
%
%
We then use KKT optimality conditions, which take into
account the constraints on the distortion spectra, and
apply well-known results on the asymptotic eigenvalue distribution,
in order to obtain a characterization of the distortion
spectra. The resulting distortion spectra are specified
via two Lagrangian parameters, which control the trade-off between
the side distortion, the central distortion, and the coding rate.




In the second part of this paper,
we propose an efficient {\em time-domain} realization for the symmetric colored
QG-MD RDF.
In~\cite{ostergaard:2009}, it was shown that Ozarow's white Gaussian
MD RDF can be achieved by noise-shaped coding based on dithered Delta-Sigma
quantization (DSQ), followed by memoryless entropy coding. Furthermore, by exploiting the fact
that Ozarow's test channel becomes asymptotically optimal for
stationary sources in the high-rate regime~\cite{zamir:2000}, it was
shown in \cite{ostergaard:2009} that, at high resolution, the
stationary MD RDF is achievable by DSQ and \emph{joint} entropy
coding.


Our time-domain approach extends the scheme of \cite{ostergaard:2009}
to the colored case at a general resolution.
We show that
the symmetric colored QG MD RDF can be achieved by oversampled noise-shaped predictive coding and \emph{memoryless} dithered quantization (in the limit of high dimensional quantization)
at all resolutions and all side-to-central distortion ratios.
We establish this result by forming a nested prediction / noise-shaping structure
containing a dithered DSQ scheme similar to~\cite{ostergaard:2009}
in the outer loop and a predictive coder per each description in
the inner loop. 
Each of the predictive coders has the structure of a differential 
pulse-code modulation (DPCM) scheme that was shown to be
optimal in the SD setting in~\cite{zamir:2008}. 
These predictive coders exploit the source memory,
and thereby minimize the coding rate and make sure that {\em memoryless
entropy coding} is optimal.\footnote{The idea of exploiting prediction in MD coding has previously been
proposed by other authors, see for example the following related
works~\cite{ingle:1995,regunthan:2000,vaishampayan:1999,nathan:2001}.
All these works faced the basic problem: Since DPCM uses prediction
from the reconstruction rather than from the source itself, and this
prediction should be reproduced at the decoder, it is not clear
which of the possible reconstructions should be used for prediction.
The present work solves this problem, using a combination of oversampling,
and Nyquist-rate side-receiver-based prediction.
}
The role of the DSQ loop is to shape the quantization noise so that
a desired trade-off between the side distortions and the central
distortion is achieved. 
At general resolutions, the optimal
noise shaping is determined by the
two-parameter solution in the first part of the paper, and it depends
upon the source spectrum. However, at high
resolutions, the optimal noise shaping becomes independent of the source
spectrum, and converges to that of a white source \cite{ostergaard:2009}:
a piece-wise constant function with a single jump discontinuity.


The paper is organized as follows.
Section~\ref{sec:prelim} provides the preliminaries.
Then, Section~\ref{sec:qrdf} proposes a differential test channel, which provides a
new interpretation of the QG MD RDF
of \cite{ozarow:1980} and \cite{chen:2009}.
The optimal distortion spectra are derived in Section~\ref{sec:spectral}.
With the test channel in mind, Section~\ref{sec:mask} presents an SD time-domain scheme, which encodes a source subject to a distortion mask.
Then, the remaining part of Section~\ref{sec:mdscheme} extends the SD time-domain scheme of Section~\ref{sec:mask} to the MD case.
Conclusions are in Section~\ref{sec:conclusion}. Longer proofs are deferred to the appendices.


\section{Preliminaries}\label{sec:prelim}

\subsection{Stationary Gaussian Processes and Spectral Decomposition}

Let
$X=\{X[n]\}_{n=0}^{\infty}$
be a zero-mean discrete-time stationary autoregressive Gaussian process with power spectral density (PSD)
$S_X$ defined as 
\begin{equation} \label{eq:PSD}
\s \triangleq \sum_{k=-\infty}^{\infty} R_X[k] e^{-j w k} \end{equation} for $\omega\in [-\pi,\pi]$, where $R_X[k] = E\{ X[n] X[n+k] \}$
is the autocorrelation function
(which is independent of $n$ due to stationarity).
We assume that $S_X$ obeys the Paley-Wiener conditions \cite{van_trees:1968},
hence it has a positive entropy-power $0<P_e(S_X)<\infty$.
Recall that the entropy power of
a spectrum $S_X$ is given by\footnote
{
For a general stationary process $X$, the entropy power is defined as
$P_e(X) \triangleq \frac{1}{2\pi e}e^{2 \bar{h}(X)}$,
where $\bar{h}(X)$ is the entropy rate.
In the stationary Gaussian case,
$\bar{h}(X)=\frac{1}{2}\log(2\pi e) + \frac{1}{4\pi}\int_{-\pi}^{\pi}
\log(S_X(e^{j\omega})) d\omega$,
which implies \eqref{eq:entropy_power}.
}
\begin{equation}\label{eq:entropy_power}
P_e(S_X) = \exp\bigg( \frac{1}{2\pi}\int_{-\pi}^{\pi}{\log\big(\s\big)} \, d\omega \bigg),
\end{equation}
where here and onwards all logarithms are taken to the natural base
unless explicitly stated otherwise.
The source may be represented in the time-domain by
\begin{equation}\label{eq:Xn}
X[n] = \sum_{k=1}^\infty a_k X[n-k] + I[n],
\end{equation}
or in the $z$-domain by
\begin{equation}
X(z) = \frac{1}{1-A(z)}I(z),
\end{equation}
where  $\{I[n]\}_{n=0}^{\infty}$ is i.i.d.\ zero-mean Gaussian and where
\begin{equation}\label{eq:az}
A(z) \triangleq \sum_{i=1}^{\infty}a_kz^{-k}
\end{equation}
is the \emph{optimal predictor}\footnote{We use the term \emph{optimal
    predictor} to denote the unique filter, which when used to predict
  the source from its infinite past, minimizes the variance of the
  prediction error. If the source is Gaussian, the prediction error,
  i.e., $\{I[n]\}$, is a white Gaussian process~\cite{makhoul:1975}.}
associated with $S_X$.
Using this notation, a spectrum $S_X$ has a spectral decomposition~\cite{makhoul:1975}:
\beq{spectral_decomposition}
\s = \left. \frac{P_e(S_X)}{
(1-A(z))
(1-{A}^*\left(\frac{1}{z^*}\right)) }\right|_{z=e^{j\omega}}. \eeq


\subsection{Symmetric MD Coding}
%
%
Consider encoding an $N$-block $\bX = X[0],\ldots,X[N-1]$,
taken from a discrete-time continuous-valued stationary 
source $X$.  
A rate-$(R_1,R_2)$ two-description coding scheme 
consists of two encoders and three decoders.
It is characterized by a 5-tuple
$(f_1,f_2,g_1,g_2,g_c)$,
where 
$f_i: \mathbb{R}^N \rightarrow \{ 1,\ldots, 2^{N R_i} \}$,
for $i=1,2$, are the two encoding functions;
$g_i: \{ 1,\ldots, 2^{N R_i} \} \rightarrow \mathbb{R}^N$
for $i=1,2$, are the two side decoding functions;
and 
$g_c: \{ 1,\ldots, 2^{N R_1} \} \times \{ 1,\ldots, 2^{N R_2} \} 
\rightarrow \mathbb{R}^N$
is the central (or joint) decoding function.
The resulting three reconstruction $N$-blocks are
$\hat{\bX}_i = g_i(f_i(\bX))$, for $i=1,2$,
and 
$\hat{\bX}_C = g_c(f_1(\bX), f_2(\bX))$.
%


A rate pair $({R}_1,{R}_2)$ is said to be achievable with respect to a
mean-squared distortion triplet $({D}_1,{D}_2,{D}_C)$, 
if for a sufficiently large block length $N$, there exists 
a rate-$({R}_1,{R}_2)$ coding scheme
$(f_1,f_2,g_1,g_2,g_c)$,
%
such that the side and central distortions satisfy
\begin{equation}
\frac{1}{N}\sum_{n=0}^{N-1}{E}[(X[n] -
\hat{X}_i[n] )^2 ] \leq {D}_i, \quad i=1, 2,
\end{equation}
and
\begin{equation}
\frac{1}{N}\sum_{n=0}^{N-1}\mathbb{E}[(X[n] -
\hat{X}_C[n] )^2 ] \leq D_C.
\end{equation}


In this work, we are interested in the symmetric case,
where $R_1=R_2 \triangleq R$ and $D_1=D_2\triangleq D_S$. 
The {\em symmetric MD RDF} of the source $X$ is defined as the minimum 
rate $R$ per description, which is achievable with respect to the 
distortion pair $(D_S,D_C)$.


\begin{remark}
In Section \ref{sec:mdscheme},
we shall consider a randomized (dithered) coding scheme,
where the encoding and decoding functions $(f_1,f_2,g_1,g_2,g_c)$
depend also on a common randomness.
Although such randomization cannot improve the achievable rate region, \cite{chen:2009},
we shall use it to simplify the analysis of
a specific (lattice-based) coding scheme.
\end{remark}



\subsection{The Quadratic-Gaussian Case}
When the source is white-Gaussian with
variance $\sigma_X^2$,
%
the symmetric MD RDF
is given by Ozarow \cite{ozarow:1980}:
%
%
\begin{equation}
\label{Ozarow_rate}
R_{white}(\sigma_X^2,D_C,D_S) \triangleq \frac{1}{4}
\!\log\!\bigg(\!
\frac{\sigma_X^2(\sigma_X^2-D_C)^2}{4D_C(D_S-D_C)(\sigma_X^2-D_S)} \!\bigg)
\end{equation}
for all {\em non-degenerate distortion pairs},
which are the non-negative ones that satisfy:
\begin{subequations}
\label{non-degenerate}
\begin{eqnarray}
\label{non-degenerate1}
D_S &\leq& \sigma_X^2  
\\
\label{non-degenerate2}
D_C &\geq& 2D_S - \sigma_X^2  
\\
\label{non-degenerate3}
D_C &\leq& \left( \frac{2}{D_S} - \frac{1}{\sigma_X^2} \right)^{-1}
\triangleq D_{C {\rm max}} .
\end{eqnarray}
\end{subequations}
%
%
For the maximum distortion point, we define separately 
$R_{white}(\sigma_X^2,\sigma_X^2,\sigma_X^2)=0$.\footnote{One may verify that this is indeed the limit of \eqref{Ozarow_rate} as $D_C\rightarrow \sigma_X^2$, $D_S\rightarrow \sigma_X^2$. This holds since $D_C$ can be sandwiched using \eqref{non-degenerate2} and \eqref{non-degenerate3}, yielding:
\[ \lim_{D_C\rightarrow\sigma_X^2} \frac{\sigma_X^2-D_C}{\sigma_X^2-D_S} = \lim_{D_C\rightarrow\sigma_X^2} \frac{\sigma_X^2-D_C}{D_S-D_C} = 2. \]}
Note that condition \eqref{non-degenerate2} applies only to high distortions
($\sigma_X^2/2 < D_S < \sigma_X^2$), and it is void otherwise;
see \cite{chen:2009}.
In the non-degenerate distortion regime,
$R_{white}(\sigma_X^2,D_C,D_S)$ monotonically decreases
with $D_C$, and reaches a saturation for $D_C = D_{C {\rm max}}$
of \eqref{non-degenerate3};
i.e.,
$R_{white}(\sigma_X^2,D_C,D_S)
=
R_{white}(\sigma_X^2,D_{C {\rm max}},D_S)$
for all
$D_C > D_{C {\rm max}}$.  
Similarly, the RDF decreases with $D_S$ for fixed $D_C$ until it saturates for a maximal $D_S$ according to \eqref{non-degenerate1} or \eqref{non-degenerate2}.


For a colored Gaussian source,
Theorem 4 of \cite{chen:2009} reduces in the symmetric case to:
\begin{equation}
\label{ChenTianDiggaviRate}
\begin{split}
&R(S_X,D_C,D_S) = \min_{\{\Ds\},\{\Dc\}} \\
&\bigg\{ \frac{1}{2 \pi} \int_{-\pi}^{\pi}
R_{\rm white} \left( S_X(\ej), \Dc, \Ds \right) dw \bigg\}, 
\end{split}
\end{equation}
where
$R_{white}$ is given in \eqref{Ozarow_rate},
and where the minimization is carried out over all
{\em distortion spectra} $\Ds$ and $\Dc$ satisfying
\beqn{ChenTianDiggaviRate2}
\inthalf{\Ds} & \leq & D_S
\nonumber \\
\inthalf{\Dc} & \leq & D_C .
\eeqn
%
%
It follows from the properties of $R_{white}(\sigma_X^2,D_C,D_S)$ above,
that 
one may restrict the minimization in \eqref{ChenTianDiggaviRate} 
to spectra that are everywhere
non-degenerate as in \eqref{non-degenerate}; i.e.,\footnote{
Since if at some frequency $\Dc$ saturates, 
we may reduce the contribution of that frequency to the total distortion
without increasing its contribution to the total rate.
}
\beqn{non-degenerate_spectra}
\Ds &\leq& S_X(\ej)  \nonumber \\
\Dc &\geq& 2\Ds - S_X(\ej) \nonumber \\
\Dc &\leq& \left( \frac{2}{\Ds} - \frac{1}{S_X(\ej)} \right)^{-1}
\eeqn
for all $\omega$.
%
%

%


%
%
%


\subsection{Additional Notation}
For $x$ real or complex, $\sqrt[n]{x}$ has $n$ roots.
For $n=2$ and $0\leq x\in\mathbb{R}$ we define $\sqr{x} \triangleq |\sqrt{x}|$, i.e., it is always non-negative. For $0>x\in\mathbb{R}$ we define $\sqr{x} \triangleq i|\sqr{-x}|$, i.e., we take the principal complex root. For $n=3$ and $x\in \mathbb{R}$ we let $\sqrr{x} \triangleq sign(x)|\sqrr{|x|}|$ denote the unique real cubic root of $x$, e.g., $\sqrr{-8} = -2$. If $x\in\mathbb{C}$ and $imag(x)\neq 0$, we let $\sqrr{x}$ denote the principal complex root, i.e., it has a positive imaginary part.
We use the notation $\xi_i^\Xi$ to indicate the $i$th root of the
function $\Xi$. If $\varphi$ is a function of $\zeta$, we use the
notation $\varphi|_{\zeta = \lambda}$ to indicate that the function
$\varphi$ is evaluated at the point $\zeta=\lambda$. If $\Phi$ is a
matrix, we use $|\Phi|$ to denote the determinant of $\Phi$.


\section{Differential Form of Ozarow's Test Channel}
%
\label{sec:qrdf}
In this section we re-state known results about the QG MD achievable rate in the symmetric case, in order to gain some insight and prepare the ground for what follows.
The exposition is made simpler by starting with a white source at the high resolution
regime.  We then proceed to general resolution and to colored sources.



\subsection{White Source, High Resolution}
\begin{figure}[t]
\begin{center}
\psfrag{X}{$\scriptstyle X$}
\psfrag{X1}{$\scriptstyle \hat{X}_1$}
\psfrag{X2}{$\scriptstyle \hat{X}_2$}
\psfrag{Xc}{$\scriptstyle \hat{X}_C$}
\psfrag{Z+Z}{$\scriptstyle Z_+ + Z_-$}
\psfrag{Z-Z}{$\scriptstyle Z_+ - Z_-$}
\psfrag{1/2}{$\scriptstyle \ \frac{1}{2}$}
\includegraphics[width=6cm]{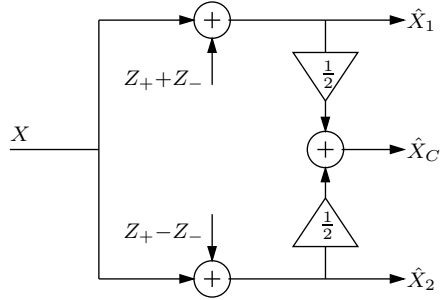}
\caption{A differential form of Ozarow's test channel for symmetric rates and distortions at high resolution conditions.}
\label{Ozarow_high}
\end{center}
\end{figure}

Let us first define the following rate expression
\beq{Ozarow_rate_HR}
R_{white,HR}(\sigma_X^2,D_C,D_S) \triangleq
\frac{1}{2}\log\left(\frac{\sigma_X^2}{2\sqrt{ (D_S-D_C) D_C }} \right),
\eeq
for
\begin{equation}
\label{non-degenerate_HR}
D_C \leq \frac{D_S}{2} .  
\end{equation}
%
In the high-resolution limit, where $\sigma_X^2\gg \max\{D_C,D_S\}$,
the symmetric QG MD RDF (\ref{Ozarow_rate})
converges to \eqref{Ozarow_rate_HR},
in the sense that the difference
$R_{white}(\sigma_X^2,D_C,D_S) - R_{white,HR}(\sigma_X^2,D_C,D_S)$
goes to zero, as $\sigma_X^2$ goes to infinity for fixed distortion levels (see \cite{zamir:1999}).
In this limit, the maximal central distortion $D_{C {\rm max}}$ \eqref{non-degenerate3} approaches $D_S/2$.
If the central decoder were to linearly combine two descriptions
of mutually independent errors of variance $D_S$ each, it would
achieve exactly this maximal distortion. 
To further reduce $D_C$ below $D_S/2$, the individual description
errors must be \emph{negatively correlated}.
Indeed, in Ozarow's test channel (see \cite{ozarow:1980}),
the relation between the side and central noises can be
explained by the side noises having a correlation matrix:
\beq{corr1} \Phi = D_S \left[\begin{array}{cc}
            1 & \rho  \\
            \rho & 1 \\
        \end{array} \right],  \eeq
where the correlation coefficient
\[ \rho=-\frac{D_S-2D_C}{D_S} \]
is negative for $D_C < D_S/2$.  
With this notation,
\eqref{Ozarow_rate_HR} becomes:
\begin{align}\label{Ozarow_rate_det}
R_{white,HR}&(\sigma_X^2,D_C,D_S)
= \frac{1}{2}\log\left(\frac{\sigma_X^2}{\sqrt{|\Phi|}}\right) \\
&=
\frac{1}{2}\log\left(\frac{\sigma_X^2}{D_S}\right) +
\frac{1}{2}\log\left(\frac{1}{\sqrt{1-\rho^2}} \right) \\
&= 
\frac{1}{2}\log\left(\frac{\sigma_X^2}{D_S}\right) + \frac{1}{2} \delta_{HR},
\end{align}
where
$|\Phi|$ denotes the absolute determinant of the matrix $\Phi$,
and
$\delta_{HR}
\triangleq
- \frac{1}{2}\log (1-\rho^2)$
is the high-resolution excess
rate \cite{zamir:1999}.
Still in the high-resolution case, we take
another step: Without loss of generality, we can represent the
correlated noises as the sum of two mutually independent noises, one
is added to both branches while the other is added to one branch and
subtracted from the other, as depicted in \figref{Ozarow_high}. In the
figure, $\hat{X}_1$ and $\hat{X}_2$ refer to the two side
descriptions, and $\hat{X}_C$ refers to the central reconstruction.
Note that the averaging eliminates $Z_-$ from the central
reconstruction $\hat{X}_C$. If we denote the variances of the noises $Z_+$ and
$Z_-$ as $\Theta_+$ and $\Theta_-$, respectively, then we can
re-write \eqref{corr1} as:
\beq{corr2} \Phi =
\left[\begin{array}{lr}
            \Theta_++\Theta_- \ & \ \Theta_+-\Theta_- \\
            \Theta_+-\Theta_- \ & \ \Theta_++\Theta_- \\

        \end{array} \right], \eeq
with
        \beq{det_phi} |\Phi| = 4 \Theta_+\Theta_-. \eeq
Further, the distortions are related to these variances by:
\begin{subequations}
\label{distortions_high}
\beqn{distortions_high1}
D_S &=& \Theta_+ + \Theta_-,  
\eeqn
\beqn{distortions_high2}
D_C &=& \Theta_+.
\eeqn
\end{subequations}
%
%
The condition \eqref{non-degenerate_HR} for non-degenerate distortions
in the high-resolution regime now becomes
\beq{theta_region_HR}
0 \leq  \Theta_+ \leq \Theta_-    
\eeq
(which, on account of the noise covariance matrix \eqref{corr2},
confirms that only non-positive correlations are considered).

%

The following proposition expresses the high-resolution optimal rate
\eqref{Ozarow_rate_HR}, based on the
relations above.  


\begin{proposition}[Parametric representation of $R_{white,HR}(\sigma_X^2,D_S,D_C)$]
\label{prop_white_HR}
For a white-Gaussian source of variance $\sigma_X^2$,
and non-degenerate distortion pairs \eqref{non-degenerate_HR},
the high-resolution symmetric MD RDF \eqref{Ozarow_rate_HR}
is given by
\beq{rate_det}
R_{white,HR}(\sigma_X^2,D_S,D_C)
=
\frac{1}{2} \log \left(\frac {\sigma_X^2}{2 \sqrt{\Theta_+ \Theta_-}}\right)
\eeq
where  
$\Theta_+$ and $\Theta_-$ are determined by $D_S$ and $D_C$
via \eqref{distortions_high}, and they satisfy \eqref{theta_region_HR}.
%
\end{proposition}


\begin{figure}[t]
\begin{center}
\psfrag{X}{$\scriptstyle X$}
\psfrag{X1}{$\scriptstyle \hat{X}_1$}
\psfrag{X2}{$\scriptstyle \hat{X}_2$}
\psfrag{Xc}{$\scriptstyle \hat{X}_C$}
\psfrag{Z+Z}{$\scriptstyle Z_+ + Z_-$}
\psfrag{Z-Z}{$\scriptstyle Z_+ - Z_-$}
\psfrag{1/2}{$\scriptstyle \ \frac{1}{2}$}
\psfrag{as}{$\scriptstyle \alpha_S$}
\psfrag{ac}{$\scriptstyle \alpha_C$}
\psfrag{V1}{$\scriptstyle V_1$}
\psfrag{V2}{$\scriptstyle V_2$}
\psfrag{Vc}{$\scriptstyle V_C$}
\psfrag{U}{$\scriptstyle U$}
\includegraphics{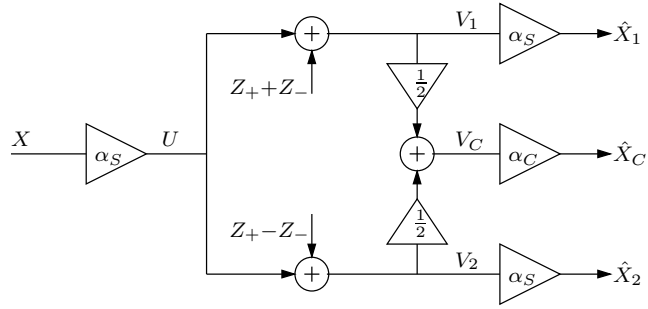}
\caption{A differential form of Ozarow's test channel with pre- and post factors, for coding at general resolution.}
\label{Ozarow}
\end{center}
\end{figure}


\subsection{White Source, General Resolution}
%
%
We now generalize our view to all distortion levels, where the
symmetric MD RDF is given by Ozarow's formula \eqref{Ozarow_rate}.
A similar correlated-noises
model to \eqref{corr1} can be obtained by expressing $\rho$ in a
rather complicated form in terms of $D_S$, $D_C$ and $\sigma_X^2$ 
\cite{ozarow:1980}.
However, we can greatly simplify such an
expression by proper use of pre- and post-factors, as depicted in
\figref{Ozarow} by $\alpha_S$ and $\alpha_C$.

In the SD case, 
it is convenient to make the pre factor
equal to the post factor \cite{gallager:1968},\cite{zamir:1996}.
However, this is
generally not possible in MD coding, because
these factors are tuned according to the signal-to-distortion
ratio, which is 
different for the side and central reconstructions.
We match the pre-factor to the {\em side} signal-to-distortion ratio;
i.e., take it to be equal to the side post-factor.
While this choice seems arbitrary, it will
prove useful in the sequel. 


The pre- and post factors are given by
\begin{subequations}
\label{alphas}
\begin{eqnarray}
\label{alphas1}
\alpha_S & \triangleq & \sqrt {\frac
{\sigma_X^2 - \Theta_+ - \Theta_-}
{\sigma_X^2}}, 
\\
\label{alphas2}
\alpha_C & \triangleq & \frac {\alpha_S \sigma_X^2}{\alpha_S^2 \sigma_X^2
+ \Theta_+} = \sqrt{\frac{\sigma_X^2(\sigma_X^2 - \Theta_+ -
\Theta_-)}{(\sigma_X^2 - \Theta_-)^2}}. \hphantom{aa}
\end{eqnarray}
\end{subequations}
With these values, noting that between $U$ and $\{V_1, V_2, V_C\}$
in Figure~\ref{Ozarow} we have exactly the
additive-noise test channel of Figure~\ref{Ozarow_high},
the resulting side and central distortions are
\begin{subequations}
\label{distortions}
\begin{eqnarray}
\label{distortions1}
D_S &=& \Theta_+ +
\Theta_-,  
\\
\label{distortions2}
D_C &=& \frac{\sigma_X^2 \Theta_+}{\sigma_X^2 - \Theta_-},
\end{eqnarray}
\end{subequations}
where as before $\Theta_+$ and $\Theta_-$ are the variances of the
noises $Z_+$ and $Z_-$ in \figref{Ozarow}.
We obtain the $\alpha_S$ and $D_S$ above from the well known (symmetric) forward test channel realization of the QG RDF in the SD case (see \cite{gallager:1968}, \cite{zamir:1996}),
where the equivalent noise variances (the variances of $Z_+ + Z_-$ and $Z_+ - Z_-$) is  $\Theta_+ + \Theta_-$.
For the pre-factor $\alpha_S$ in \eqref{alphas1}, 
the optimum solution for $\alpha_C$ and $D_C$ 
in \eqref{alphas2} and \eqref{distortions2}
is then given by Wiener estimation of $X$ from the measurement 
$V_C = \alpha_S X + Z_+$.


The choice of factors we have made provides a simple parametric 
characterization for the symmetric MD RDF.



\begin{proposition}[Parametric representation of $R_{white}(\sigma_X^2,D_S,D_C)$]
\label{prop_white}
%
For a white Gaussian source of variance $\sigma_X^2$,
and non-degenerate distortion pairs \eqref{non-degenerate}, 
the symmetric MD RDF \eqref{Ozarow_rate}
can be written as
\beq{rate_det2}
R_{white}(\sigma_X^2,D_S,D_C)
=
\frac{1}{2} \log \left(\frac {\sigma_X^2}{2 \sqrt{\Theta_+ \Theta_-}}\right) ,
\eeq
where  
$\Theta_+$ and $\Theta_-$ are determined by $D_S$ and $D_C$ 
via \eqref{distortions}.
Furthermore,
the non-degenerate distortion condition \eqref{non-degenerate}
can be written as
\beq{theta_region}
0 \leq  \Theta_+ \leq \Theta_-  \leq \frac{\sigma_X^2}{2} ,
\eeq
%
where the maximum central-distortion (equality in \eqref{non-degenerate3})
occurs at $\Theta_+ = \Theta_-$.
\end{proposition}


\begin{IEEEproof}
See Appendix~\ref{app:prop_white}.
\end{IEEEproof}

%


Note that when 
$\sigma_X^2 \gg \Theta_+ + \Theta_-$,
the pre- and post-factors \eqref{alphas} approach 1,
and (\ref{distortions}) reduces to (\ref{distortions_high});
so we are back in the high-resolution case of the previous section.
%


\subsection{Stationary Source}

\begin{figure}[t]
\begin{center}
\psfrag{X}{$\scriptstyle X[n]$}
\psfrag{X1}{$\scriptstyle \hat{X}_1[n]$}
\psfrag{X2}{$\scriptstyle \hat{X}_2[n]$}
\psfrag{Xc}{$\scriptstyle \hat{X}_C[n]$}
\psfrag{Z+Z}{\raisebox{-3mm}{$\scriptstyle Z_+[n] + Z_-[n]$}}
\psfrag{Z-Z}{\raisebox{3.5mm}{$\scriptstyle Z_+[n] - Z_-[n]$}}
\psfrag{1/2}{$\scriptstyle \ \frac{1}{2}$}
\psfrag{as}{$\scriptstyle \alpha_S$}
\psfrag{ac}{$\scriptstyle \alpha_C$}
\psfrag{V1}{$\hspace{-2mm} \scriptstyle V_1[n]$}
\psfrag{V2}{$\hspace{-2mm} \scriptstyle V_2[n]$}
\psfrag{Vc}{$\hspace{-2mm} \scriptstyle V_C[n]$}
\psfrag{U}{$\scriptstyle U$}
\psfrag{X}{$\scriptstyle X[n]$}
\psfrag{U}{$\hspace{-1mm} \scriptstyle U[n]$}
\psfrag{B}{$\scriptstyle B[n]$}
\psfrag{D}{$\scriptstyle D[n]$}
\psfrag{Y}{$\scriptstyle Y[n]$}
\psfrag{V}{$\scriptstyle V[n]$}
\psfrag{Z}{$\scriptstyle Z[n]$}
\psfrag{E}{$\scriptstyle E[n]$}
\psfrag{Et}{$\scriptstyle \tilde{E}[n]$}
\psfrag{Xh}{$\scriptstyle \hat{X}[n]$}
\psfrag{F(z)}{$\scriptstyle F(\ej)$}
\psfrag{A(z)}{$\scriptstyle \ A(z)$}
\psfrag{F*(z)}{$\hspace{-2mm} \scriptstyle \ F^{*}(\ej)$}
\psfrag{G*(z)}{$\hspace{-2mm} \scriptstyle \ G^{*}(\ej)$}
\psfrag{C(z)-1}{$\scriptstyle \ \ C(z)$}
\includegraphics[width=0.95\columnwidth]{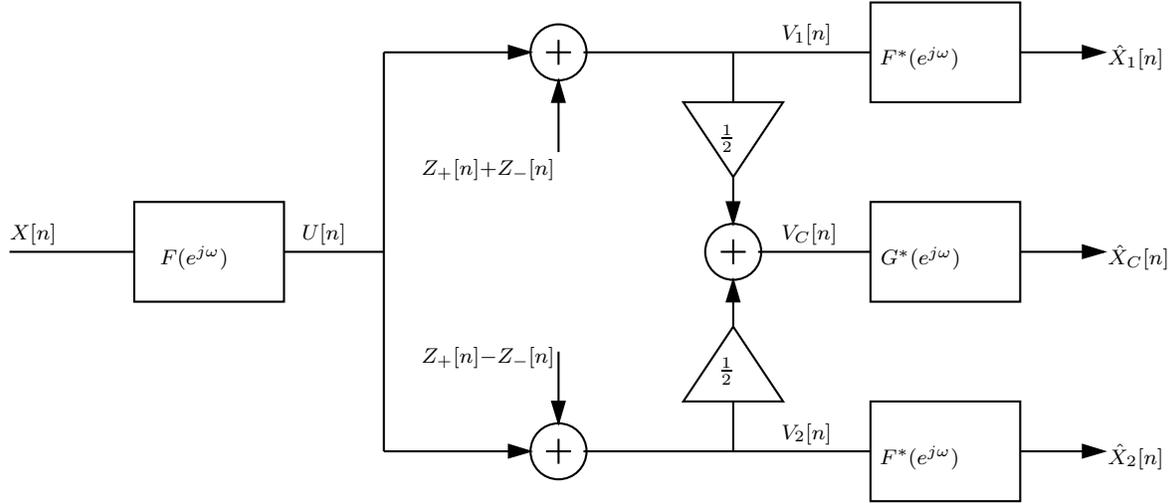}
\caption{Pre/post-filtered differential test channel for a colored Gaussian source.}
\label{Ozarow_colored}
\end{center}
\end{figure}


We turn to a general colored Gaussian source, 
whose symmetric MD RDF \eqref{ChenTianDiggaviRate} was derived
by Chen et al. \cite{chen:2009}.
We can re-write this formula in the spirit of the above exposition.
To that end, we present in \figref{Ozarow_colored} a test channel,
which may be seen as a colored version of the channel of \figref{Ozarow}.
In this channel, the noise processes $Z_+$ and $Z_-$ are stationary Gaussian,
with spectra $\{\Theta_+(e^{j2\omega})\}$ and $\{\Theta_{-}(e^{j2\omega})\}$,
respectively.
The factors $\alpha_S$ and $\alpha_C$ of \figref{Ozarow} are replaced by
linear time-invariant filters, with frequency response $F(\ej)$ and $G(\ej)$,
respectively.
At each frequency $\omega$, the 5-tuple
$\{ \sx, \Tp, \Tm, F(\ej), G(\ej) \}$
is intra-related as
$\{ \sigma_X^2, \Theta_+, \Theta_-, \alpha_S, \alpha_C \}$
in the optimal scalar test channel of \figref{Ozarow}.
That is, in view of \eqref{alphas},
the filters satisfy:\footnote{
Without loss of generality, one may take all filters to have real frequency response.
However, the more general complex form allows more flexibility;
e.g. one of the filters may be made causal.
}
\begin{subequations}
\label{filters}
\begin{eqnarray}
\label{filters1}
|F(\ej)|^2 =& \frac{\sx -\Tp - \Tm}{\sx} \\
\label{filters2}
|G(\ej)|^2 =& \frac{\sx(\sx - \Tp - \Tm)}{(\sx - \Tm)^2} 
\end{eqnarray}
where the phases of the filters are equal, but otherwise arbitrary:
\begin{equation}
\label{filters3}
\angle{F(\ej)} = \angle{G(\ej)} .
\end{equation}
\end{subequations}


\rem{
It would be convenient to define the equivalent up-sampled noise
spectrum:
\beq{theta}
\tTheta(\ej) = \left\{
\begin{array}{ll}
            2\Theta_+\left(e^{j2\omega}\right), & \hbox{$|\omega| < \frac{\pi}{2}$}, \\
            2\Theta_-\left(e^{j2 \left(\omega-\frac{\pi}{2}\right)}\right), & \hbox{$\frac{\pi}{2} < \omega \leq \pi$}, \\
            2\Theta_-\left(e^{j2 \left(\omega+\frac{\pi}{2}\right)}\right), & \hbox{$-\pi \leq \omega < -\frac{\pi}{2}$},
        \end{array} \right.
        \eeq
where the lowpass and highpass spectra of $\tilde{\Theta}$ are
formed by $\{\Theta_+(e^{j2\omega})\}$ and $\{\Theta_{-}(e^{j2\omega})\}$,
which are compressed versions (by a factor of two) of the spectra
$\Theta_+ = \{\Tp\}_{\omega=-\pi}^{\pi}$ and $\Theta_-= \{\Tm\}_{\omega=-\pi}^{\pi}$,
respectively.
It is not hard to verify that the entropy power \eqref{eq:entropy_power}
of this spectrum is given by
\beq{theta_entropy_power}
P_e(\tTheta)
=
\exp \left(
\inthalf{ \log (2\sqrt{\Tp\Tm}) }
\right) .
\eeq
}

The following proposition gives a parametric expression for 
the symmetric colored QG MD RDF \eqref{ChenTianDiggaviRate},
in terms of entropy powers: 

\begin{proposition}[Parametric presentation of $R(S_X,D_C,D_S)$]
\label{prop_Chen}
For a stationary Gaussian source of spectrum $S_X(\ej)$,
the symmetric MD RDF \eqref{ChenTianDiggaviRate}
is parametrically given by
\begin{equation}
\label{ParametricChenTianDiggaviRate}
\begin{split}
&R(S_X,D_C,D_S)  \\
&=
\min_{\{\Tp\},\{\Tm\}}
\frac{1}{2} \log \left( \frac{ P_e(S_X) }{ 2 \sqrt{P_e(\Theta_+)P_e(\Theta_-)}} \right),
\end{split}
\end{equation}
where 
the minimization is taken over all $\Tp$ and $\Tm$
satisfying the non-degenerate condition:
%
\beq{theta_region_spectra}
0 \leq \Tp \leq \Tm \leq \frac{S_X(\ej)}{2}, \forall \omega ,
\eeq
and total distortion constraints:
\begin{subequations}
\label{spectra_constraints}
\begin{eqnarray}
\label{spectra_constraints1}
\inthalf{ \underbrace{[ \Tp+\Tm]}_{D_S(e^{j\omega})} }  &\leq& D_S
\\
\label{spectra_constraints2}
\inthalf{ \underbrace{ \frac{\sx\Tp}{\sx-\Tm}}_{D_C(e^{j\omega})} } &\leq& D_C .
\end{eqnarray}
\end{subequations}
%
\end{proposition}

\begin{figure}[t]
\begin{center}
\psfrag{sx2}{\raisebox{-1mm}{\hspace{-2.5mm}$\frac{\sx}{2}$}}
\psfrag{O+}{$\Tp$}
\psfrag{O-}{$\Tm$}
\psfrag{Ob}{Boundary region}
\psfrag{O}{Region of support}
\includegraphics{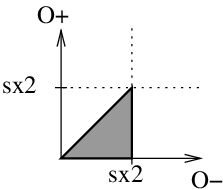}
\caption{The triangular support region of $\Tm$ and $\Tp$
as given by~(\ref{theta_region_spectra}).}
\label{fig:boundary}
\end{center}
\end{figure}

The per-frequency triangular support region
\eqref{theta_region_spectra} is
shown in Fig.~\ref{fig:boundary}.
Note that the arguments of the integrals in \eqref{spectra_constraints}
amount to the side and central {\em distortion spectral densities}
of \eqref{ChenTianDiggaviRate2}.


%
%


\begin{IEEEproof}
By Proposition~\ref{prop_white}, the minimization in \eqref{ChenTianDiggaviRate}
is equivalent to
%
\beq{explicit_minimization}  
\min_{\{\Tp\},\{\Tm\}} \inthalf{\frac{1}{2} \log \frac{S_X(\ej)}{2\sqrt{\Tp\Tm}}} ,
\eeq
where the minimization is taken over all noise spectra 
$\Tp$ and $\Tm$ satisfying
\eqref{spectra_constraints} and \eqref{theta_region_spectra}.
On account of \eqref{eq:entropy_power}, 
this completes the proof.
\end{IEEEproof}


%
In the next section, we shall provide an explicit solution for the optimal 
noise spectra $\Tp$ and $\Tm$ in Proposition~\ref{prop_Chen}, and hence
for the optimal distortion spectra $\{\Ds\}$ and $\{\Dc\}$
in \eqref{ChenTianDiggaviRate2}.


\begin{remark} 
As we shall see in the next section,
in the high resolution limit the optimal spectra $\Tp$,
$\Tm$ become flat, 
so the RDF becomes that of Proposition~\ref{prop_white_HR} with the source variance
$\sigma_X^2$ replaced by the entropy-power $P_e(S_X)$,
in accordance with \cite{Dragotti:2002}, \cite{ostergaard:2009}.
\rem{Furthermore, the same noise spectrum is optimal for a white
source at a {\em general} resolution, provided that 
$\Theta_+$ and $\Theta_-$ are determined through 
\eqref{distortions},
in accordance with Proposition~\ref{prop_white}
and \cite{ostergaard:2009}.} 
\end{remark}


\begin{remark}\label{rem:epsilon}
If $X$ does not satisfy the Paley-Wiener condition, i.e. $P_e(S_X)=0$, 
then the rate expression 
\eqref{ParametricChenTianDiggaviRate}
is not well defined.
In this case, we may use the following:
For any $\epsilon>0$, let
$S_{X_\epsilon}(e^{j\omega}) = \max (S_X(e^{j\omega}) , \epsilon), \forall \omega$,
and
$D_\epsilon = \frac{1}{2\pi}\int_{-\pi}^{\pi} \max(0,\epsilon - S_X(e^{j\omega}))  d\omega$.
Then there exists some $\epsilon>0$ such that Proposition~\ref{prop_Chen} holds with
$S_X$, $D_S$, and $D_C$ replaced by
$S_{X_\epsilon}, D_S + D_\epsilon$, and $D_C+D_\epsilon$, respectively.
\end{remark}

%

\section{Optimal Distortion Spectra with Two Parameters}
\label{sec:spectral}
In this section we derive a simple spectral domain characterization of the
optimal distortion spectra, thus of the rate-distortion function, in terms of two Lagrange parameters.
From Proposition~\ref{prop_Chen} and in particular
from~(\ref{explicit_minimization}), it may be noticed that
finding the RDF is equivalent to finding a pair of noise spectra $\{\Theta_+(e^{j\omega})\}_{\omega=-\pi}^{\pi}$ and $\{\Theta_-(e^{j\omega})\}_{\omega=-\pi}^{\pi}$, which minimizes the description rate $R$
subject to the two target distortion constraints $D_S$ and $D_C$, cf.~(\ref{spectra_constraints}). 
These noise spectra can in principle be found using results from
variational calculus, where one may cast the constrained
minimization problem as an isoperimetric problem,
cf.~\cite{weinstock:1974} for details. However, this direct way leads
to some technical subtleties regarding establishing neccesary
conditions for optimality when combining coupled inequality
constraints consisting of integrals in addition to per frequency
constraints. To circumvent these technicalities, we take here the
conventional approach from Information Theory: first solve the
optimization problem for a \emph{finite} vector of parallel sources (with a \emph{joint}
distortion constraint); then use that to optimize for a
\emph{correlated} source vector, by an orthonormal
(``Karhunen-Lo\`{e}ve'') transformation (KLT) into parallel sources whose variances are the eigenvalues of the source covariance matrix;
and finally apply Szeg\"o's
theorem to obtain spectral densities as the limit of the
eigenvalues when the dimension of the source vector goes to infinity, cf., \cite{berger:1971,CoverBook}. 


\subsection{$K$ parallel Gaussian sources}
Let us consider the case of $K$ independent Gaussian sources with variances $\{\eta(k) \geq 0\}$, $k=1,\ldots,K$. 
For any $\eta>0$, $\Theta_+\geq 0$, $\Theta_- \geq 0$ let
\begin{align} \label{functions_start} R(\eta,\tp,\tm) &\triangleq \frac{1}{2}
  \log \frac{\eta}{2\sqrt{\Theta_+\Theta_-}} \\ \label{eq:Dc_eta}
D_S(\eta,\tp,\tm) &\triangleq \tp + \tm  \\ \label{eq:Ds_eta}
D_C(\eta,\tp,\tm) &\triangleq \frac{\eta\tp}{\eta - \tm},  \end{align} while
for $\eta = 0$, we let
\begin{align} \label{functions_end} R(0,\tp,\tm) = D_C(0,\tp,\tm) =
  D_C(0,\tp,\tm) = 0. \end{align} 
Given a
vector of $K$ sources $\{\eta(k) \geq 0\}$, $k=1,\ldots,K$, the
problem is to minimize
\begin{align}\label{eq:K_opt1} R = \frac{1}{K}\sum_{k=1,\ldots,K:\eta(k) > 0} R(\eta(k),\tpk,\tmk), \end{align}
where the minimization is with respect to $\tpk$ and $\tmk, k=1,\dotsc, K,$ and 
subject to the following constraints:
\begin{align} \label{constraints_start}
\frac{1}{K}\sum_{k=1}^{K} D_C(\eta(k),\tpk,\tmk) & \leq D_C \\ \label{side_constraint}
\frac{1}{K}\sum_{k=1}^{K}  D_S(\eta(k),\tpk,\tmk) & \leq D_S
\end{align}
and, for any $k$,
\begin{equation} \label{eq:noise_constraint}
0 \leq \tpk \leq \tmk \leq \frac{\eta(k)}{2}.
\end{equation}

In order to take into account all the distortion constraints in
addition to the constraints on the support region of the spectra, we
apply the Karush-Kuhn-Tucker (KKT) conditions~\cite{boyd:2004} and form the Lagrangian $\tilde{J}$, which will be needed in the sequel:
\begin{equation}\label{eq:tilde_J}
\tilde J = \frac{1}{K}\sum_{k=1}^K \lk , \end{equation} where
\begin{align*} & \hspace{-5mm} \lk = \\
  & \mathcal{L}_{\lambda_1,\lambda_2}(\eta(k),\tpk,\tmk) \\
  - & \mu_{1}(k) \tpk - \mu_{2}(k) (\eta(k)/2-\tmk)  \\ 
- &  \mu_{3}(k) (\tmk-\tpk), \end{align*} and where 
\begin{equation}\label{eq:lagrangian}
\begin{split}
\l &=
R(\eta,\tp,\tm) + \lambda_1 D_C(\eta,\tp,\tm) \\
& \quad + \lambda_2 D_S (\eta,\tp,\tm).
\end{split}
\end{equation}

It turns out that there  is a threshold, which depends upon $\lambda_1, \lambda_2$ and
resembles the "water-level" in SD coding, where if the source variance
is below the threshold a zero rate is allocated. We formalize this observation with the following definition and lemma.

\begin{definition}[Support] \label{def:support} 
A level $\eta$ is said to be supported at parameters $\lambda_1, \lambda_2\geq 0$ if
\[  2(\lambda_1  + 2 \lambda_2)\eta   > 1 . \]
\end{definition}

\begin{lemma} \label{lem:sequence}
For the optimization problem in \eqref{eq:K_opt1} and any 
$\lambda_1>0,\lambda_2>0$, if $\eta(k)$ is supported at $\lambda_1, \lambda_2$ then the optimal pair of noise variances $\tpk, \tmk$ is strictly within the triangular support region \eqref{eq:noise_constraint}. Otherwise, the corresponding $k$th set of optimal noise variances satisfy $\tpk = \tmk= \eta(k)/2$ and $R(\eta(k), \tpk,\tmk) = 0$. \end{lemma}
\begin{IEEEproof}
See Appendix \ref{app:lem:sequence}.
\end{IEEEproof}

Let us now consider the case where $\eta(k)$ is supported and, thus, the optimal noise variances $\tpk$ and $\tmk$ are within the triangular region \eqref{eq:noise_constraint}. Then, since the Lagrangian is differentiable inside the triangular region, the optimal solution must be a stationary solution. 
We therefore first find the partial derivatives of the Lagrangian~(\ref{eq:lagrangian}), i.e.
\begin{align}\label{eq:diftheta+}
\frac{\partial \l}{\partial \tp} & =
-\frac{1}{4\tp} + \lambda_1 + \lambda_2\frac{\eta}{\eta-\tm} \\
\label{eq:diftheta-}
\frac{\partial \l}{\partial \tm} &=
-\frac{1}{4\tm} + \lambda_1 + \lambda_2\frac{\eta\tp}{(\eta-\tm)^2}.
\end{align}

Equating both~(\ref{eq:diftheta+}) and (\ref{eq:diftheta-}) to zero and jointly solving, yields
\begin{align}\label{eq:theta2}
\tm & = \Psi_{\lambda_1,\lambda_2}^\dagger(\eta) \\
\label{eq:theta1}
\tp & =
\frac{\eta - \Psi^\dagger_{\lambda_1,\lambda_2}(\eta)}
{4(\lambda_1+\lambda_2)\eta - 4\lambda_1\Psi^\dagger_{\lambda_1,\lambda_2}(\eta)},
\end{align}
where for a fixed pair $(\lambda_1, \lambda_2)\in \mathbb{R}^2$,
$\Psi^\dagger_{\lambda_1,\lambda_2}(\eta)$ denotes a real (and
positive) root of the third-order polynomial in $\Psi$ given by:
\begin{equation}\label{eq:psipoly1}
\begin{split}
\Psi^3 &- \frac{4\lambda_1\lambda_2\eta + 8\lambda_1^2\eta +
  \lambda_1}{4\lambda_1^2}\Psi^2 \\
&+ \frac{2\lambda_1 \eta + 2\lambda_2 \eta + 4\lambda_1\lambda_2\eta^2 + 4\lambda_1^2\eta^2}{4\lambda_1^2}\Psi -\frac{(\lambda_2+\lambda_1)\eta^2}{4\lambda_1^2}.
\end{split}
\end{equation}
Since~(\ref{eq:psipoly1}) is a real third-order polynomial, three solutions are possible (of which two might be complex conjugates).
Given a real polynomial $\Psi^3 +
a_2 \Psi^2 + a_1\Psi+a_0$, where the $a_i$'s
follow from~(\ref{eq:psipoly1}), we denote its discriminant $\Xi$ by
\begin{align}\label{eq:Xi}
\Xi &= q^2 + p^3,
\end{align}
where
\begin{equation} \label{eq:p}
\begin{split}
p &=\frac{a_1}{3} - \frac{a_2^2}{9} 
  = -\frac{1}{144\lambda_1^2} \\
&\times (-8\lambda_1\eta-16\lambda_2\eta+16\lambda_1\lambda_2\eta^2 + 16\lambda_1^2\eta^2 +16\eta^2\lambda_2^2 + 1)
\end{split}
\end{equation}
and
\begin{align}\label{eq:q}
q &= \frac{1}{6}(a_1a_2-3a_0) -\frac{a_2^3}{27} \nonumber \\
  &= -\frac{1}{1728\lambda_1^3}\big(
96\lambda_1\lambda_2\eta^2 - 48\lambda_1^2\eta^2 - 64\lambda_2^3\eta^3 - 96\lambda_1\lambda_2^2\eta^3 \nonumber \\
&\quad+ 96\lambda_2^2\eta^2  +96\lambda_1^2\lambda_2\eta^3+ 24\lambda_2\eta+64\lambda_1^3\eta^3+12\lambda_1\eta-1
\big).
\end{align}
For expressing the result we also need the angle $\phi$, defined as: \begin{equation} \label{eq:phi}
\phi =
\begin{cases}
\arctan(\sqr{-\Xi}/q), & q>0, \\
\pi + \arctan(\sqr{-\Xi}/q), & q< 0, \\
\pi/2, & q =0.
\end{cases}
\end{equation}

\begin{lemma}\label{lem:unique_sol}
For any $\lambda_1>0$ and $\lambda_2>0$.  If $2(\lambda_1  + 2
\lambda_2)\eta>1$, then the optimal set of variances $(\Theta_+(\lambda_1,\lambda_2,\eta),\Theta_-(\lambda_1,\lambda_2,\eta))$, which
minimizes $\mathcal{L}_{\lambda_1,\lambda_2}(\eta, \Theta_+(\lambda_1,\lambda_2,\eta), \Theta_-(\lambda_1,\lambda_2,\eta))$, is unique, within the interior of the triangular
support region $0\leq\Theta_+(\lambda_1,\lambda_2,\eta)\leq\Theta_-(\lambda_1,\lambda_2,\eta)\leq\eta /2$ , and given by 
\begin{equation}\label{eq:opt_tp}
\Theta_+(\lambda_1,\lambda_2,\eta) = \frac{\eta - \Theta_-(\lambda_1,\lambda_2,\eta)}{4(\lambda_1 + \lambda_2)\eta - 4\lambda_1 \Theta_-(\lambda_1,\lambda_2,\eta)},
\end{equation}
and if $\Xi\geq 0$, 
\begin{equation}\label{eq:cases_x}
\begin{split}
\Theta_-(\lambda_1,\lambda_2,\eta) 
&=
\sqrr{ q\! +\! \sqr{\Xi} } + \sqrr{ q\! -\! \sqr{\Xi} }
\\ &\quad + \frac{\eta}{3\lambda_1}(2\lambda_1 + \lambda_2)  +\frac{1}{12\lambda_1}, 
\end{split}
\end{equation}
otherwise $\Xi< 0$, and
\begin{equation}
\begin{split}
\Theta_-(\lambda_1,\lambda_2,\eta) 
&= -\sqr{|p|}(\cos(\phi/3) + \sqr{3}\sin(\phi/3)) \\
& \quad + \frac{\eta}{3\lambda_1}(2\lambda_1 + \lambda_2)
+\frac{1}{12\lambda_1}, 
\end{split}
\end{equation}
where the relationship between $(p, q, \Xi, \phi)$ and $(\lambda_1, \lambda_2, \eta)$ is given by \eqref{eq:Xi} -- \eqref{eq:phi}. 
If $2(\lambda_1  + 2
\lambda_2)\eta\leq 1$, then $\Theta_+(\lambda_1,\lambda_2,\eta) = \Theta_-(\lambda_1,\lambda_2,\eta) = \eta/2$. 
\end{lemma}

\begin{IEEEproof}
See Appendix \ref{app:lem:unique_sol}.
\end{IEEEproof}

In view of \eqref{eq:K_opt1} -- \eqref{side_constraint}, Lemma \ref{lem:unique_sol} implies the following theorem.
\begin{theorem}[Two-parameter form of the symmetric MD RDF for parallel sources] \label{thm:parallel}
Given the variances $\eta^K = (\eta(1), \dotsc, \eta(K))$ of a vector of $K$ parallel sources,
the MD RDF $R(\eta^K,D_C,D_S)$ can be written parametrically, in terms of the two parameters
$\lambda_1$ and $\lambda_2$, as
\begin{align}\label{eq:Rtheo1}
 R &= \frac{1}{K} \sum_k  \frac{1}{2}  \log \frac{\eta(k)}{2\sqrt{ \Theta_+(\lambda_1, \lambda_2, \eta(k)) \Theta_-(\lambda_1, \lambda_2, \eta(k))}}  \\ \label{eq:Dstheo1}
 D_S &= \frac{1}{K} \sum_k   \Theta_+(\lambda_1, \lambda_2, \eta(k)) +  \Theta_-(\lambda_1, \lambda_2, \eta(k)) \\ \label{eq:Dctheo1}
  D_C &= \frac{1}{K} \sum_k  \frac{\eta(k) \Theta_+(\lambda_1, \lambda_2, \eta(k))}{\eta(k) -  \Theta_-(\lambda_1, \lambda_2, \eta(k))},
\end{align}
where the functions  $\Theta_+(\cdot,\cdot,\cdot)$ and $\Theta_-(\cdot,\cdot,\cdot)$ are defined in Lemma~\ref{lem:unique_sol}.
\end{theorem}


\subsection{A Correlated Gaussian Source Vector}\label{sec:source_vector}
For any \emph{finite} $K$, let $X^K\in \mathbb{R}^K$ be a zero-mean Gaussian source vector having possibly correlated elements, s.t. $\mathbb{E}\|X^K\|^2<\infty$. Further, let $\eta(k), k=1,\ldots,K$ be the eigenvalues of the covariance matrix of $X^K$. Then, the resulting source vector $\tilde{X}^K$ obtained by applying a KLT on $X^K$ is Gaussian with independent components having variances $\{\eta(k)\geq 0\}$. The optimal noise variances $\{\tpk,\tmk\}$ for $\tilde{X}^K$ are found as above. The corresponding rate and distortions for $\tilde{X}^K$ are the same as those of $X^K$, since the KLT is unitary.

\subsection{A Stationary Gaussian Process} 
We finally tend to the case of a zero-mean stationary Gaussian process
$X$ with autocorrelation function $R_X$ and finite variance. The RDF of a process is defined as the limit of the RDF per sample of
finite vectors, and we will therefore consider the infinite-dimensional limit of the Gaussian vector of Section~\ref{sec:source_vector}. By Szeg\"o's asymptotic
eigenvalue distribution theorem (see e.g. \cite{grenander:1984}), the eigenvalues of $R_X$ approach the spectrum of the process. 
The result is stated in the following definition and proposition. 

\begin{definition} \label{def:supportf} 
The set of support frequencies $\Omega(\lambda_1,\lambda_2)$ for a fixed pair $(\lambda_1, \lambda_2)$ and source PSD $\{\sx\}$ is defined by:
\begin{equation}\label{eq:supportf} 
\Omega(\lambda_1,\lambda_2) \triangleq \{ -\pi < \omega \leq \pi : 2(\lambda_1 + 2\lambda_2)\sx > 1\}.
\end{equation}
\end{definition}

\begin{proposition}[Two-parameter form of the symmetric MD RDF for stationary processes]
\label{prop:stationary_RDF}
 Let $X$ be stationary Gaussian with finite variance and with PSD $S_X$. Then
 the symmetric quadratic two-description RDF of $X$
 is given for all non-degenerate distortion pairs by taking the lower
 envelope (in the rate axis) of the $(R,D_S,D_C)$ region formed by
 scanning all $\lambda_1 \geq 0,\lambda_2 \geq 0$, and setting:
\begin{align} \nonumber
R & = \frac{1}{4\pi} \int_{\omega = - \pi}^\pi 
 \log\left(\sx\right) \\ \label{theo:R}
&-\log\left(2\sqrt{ \Theta_+(\lambda_1, \lambda_2, \sx) \Theta_-(\lambda_1, \lambda_2, \sx)}\right)\,  d\omega
 \\ \nonumber
D_S & = \frac{1}{2\pi} \int_{\omega = - \pi}^\pi  
 \Theta_+\left(\lambda_1,\lambda_2,\sx\right)  \\  \label{theo:Ds} 
& \quad+ \Theta_-\left(\lambda_1,\lambda_2,\sx\right)  \, d\omega 
 \\ \label{theo:Dc}
D_C & = \frac{1}{2\pi} \int_{\omega = - \pi}^\pi\frac{\sx \Theta_+(\lambda_1, \lambda_2, \sx)}{\sx -  \Theta_-(\lambda_1, \lambda_2, \sx)} \, d\omega,
\end{align}
where $\Theta_+(\lambda_1,\lambda_2,\sx)$ and
$\Theta_-(\lambda_1,\lambda_2,\sx)$ are the optimal per-frequency spectral
components given by \eqref{eq:opt_tp} and \eqref{eq:cases_x}, respectively, for $\omega \in \Omega(\lambda_1,\lambda_2)$, and $\Theta_+(\lambda_1,\lambda_2,\sx)=\Theta_-(\lambda_1,\lambda_2,\sx)=\sx/2$ otherwise.
\end{proposition}

Note that the rate integral \eqref{theo:R} could be equivalently
limited to the support frequency set $\Omega(\lambda_1,\lambda_2)$ in \eqref{eq:supportf};
the distortion integrals \eqref{theo:Ds},\eqref{theo:Dc} could be similarly limited, with an additional term which is the source power outside the support frequencies.

The proof of the proposition is not included in this paper, as applying the asymptotic eigenvalue distribution theorem is a standard procedure in the IT literature, see e.g. \cite{CoverBook}. Subtleties having to do with the specific nature of the solution to the MD problem have been dealt with in \cite{chen:2009}. We do include a brief outline. In order to invoke Szeg\"o's Theorem, one needs to verify two conditions: that the eigenvalues are bounded, and that the function is continuous w.r.t. these eigenvalues. We can first assume a bounded spectrum, then the eigenvalues are also bounded. Continuity can be shown in a straightforward manner, thus Theorem~\ref{thm:parallel}  directly translates to Proposition~\ref{prop:stationary_RDF}. Then, in order to extend the result to unbounded spectra (with finite variance), one can use continuity and monotonicity arguments, combined with the fact that by Proposition~\ref{prop_Chen}, which is a restatement of \cite[Theorem 4]{chen:2009}, the MD RDF is given by the form \eqref{theo:R}-\eqref{theo:Dc} for \emph{some} choice of $\Theta_-(\lambda_1, \lambda_2, \sx)$ and $\Theta_+(\lambda_1, \lambda_2, \sx)$, even when $\sx$ is unbounded.

\subsection{High-Resolution Limit}
In this section we consider the high-resolution regime and show that
the optimal noise spectra become approximately flat as is the case for
white Gaussian sources. 

\begin{proposition}[High-Rate Cases]\label{prop:hr}
Let $X$ be stationary Gaussian with finite variance and with PSD $S_X$. Then, for any $\omega\in [-\pi;\pi]$, 
\begin{equation}\label{eq:tm}
\mathop{\lim_{\lambda_1,\lambda_2\to\infty}}_{\lambda_1/\sqrr{\lambda_2}\to 0}  \lambda_1 \Theta_-(\lambda_1,\lambda_2,\sx)  = \frac{1}{4}
\end{equation}
and
\begin{equation}\label{eq:tp}
\mathop{\lim_{\lambda_1,\lambda_2\to\infty}}_{\lambda_1/\sqrr{\lambda_2}\to 0}  (\lambda_1+\lambda_2) \Theta_+(\lambda_1,\lambda_2,\sx)  = \frac{1}{4}.
\end{equation}
\end{proposition}

\begin{IEEEproof}
See Appendix~\ref{app:proof:prop:hr}.
\end{IEEEproof}

\begin{remark}
The convergence requirement of $\lambda_1/\sqrr{\lambda_2}\to 0$ in Proposition~\ref{prop:hr} is a technicality needed in the proof. 
The limiting behavior of (\ref{eq:tm}) and (\ref{eq:tp}) can also be
observed for small $\lambda_2$ and large $\lambda_1$. This shows that under
high-resolution conditions, the optimum noise spectra are flat, independent of the source spectrum, and approximately given by $\Theta_-(\lambda_1,\lambda_2,\sx) \approx \frac{1}{4\lambda_1}$ and $\Theta_+(\lambda_1,\lambda_2,\sx) \approx \frac{1}{4(\lambda_1+\lambda_2)}$.  
\end{remark}

\subsection{Example}
\begin{figure}[th]
\psfrag{w}{\scriptsize$\omega$}
\psfrag{Sx}{\scriptsize$\sx$}
\psfrag{Theta-}{\scriptsize$\Theta_-(e^{j\omega})$}
\psfrag{Theta+Theta}{\scriptsize$\Theta_+(e^{j\omega})$}
\psfrag{Dc}{\scriptsize$D_C(e^{j\omega})$}
\psfrag{Ds}{\scriptsize$D_S(e^{j\omega})$}
\psfrag{Rw}{\scriptsize$R(e^{j\omega})$}
\psfrag{pi/3}{\scriptsize\hspace{0.5mm}$\frac{\pi}3$}
\psfrag{2/3 pi}{\scriptsize\hspace{0.5mm}$\frac{2\pi}3$}
\psfrag{pi}{\scriptsize$\pi$}
\begin{center}
\includegraphics[width=8cm]{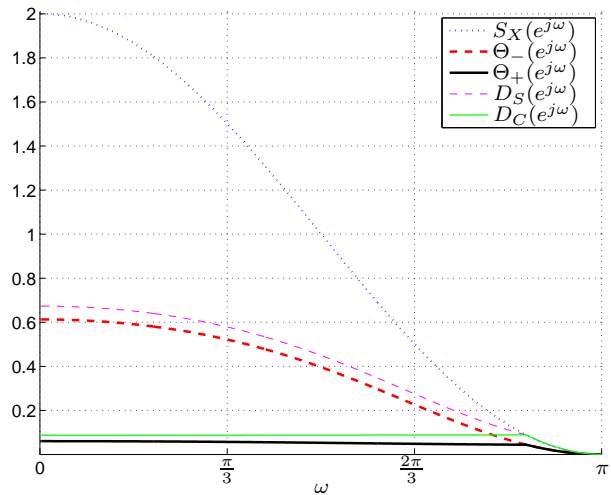}
\caption{Source, noise, and distortion spectra for $0\leq |\omega|<\pi$.}
\label{fig:rdplots}
\end{center}
\end{figure}

To elucidate the behavior of the noise spectra $\Tm\triangleq\Theta_-(\lambda_1,\lambda_2,\sx) $ and $\Tp \triangleq \Theta_+(\lambda_1,\lambda_2,\sx)$ as a function of $\lambda_1$ and $\lambda_2$, we present the following example. Let $X$ be a stationary Gaussian process with a positive and monotonically decreasing spectrum given by 
\begin{equation}\label{eq:sx}
\sx = \cos(\omega)+1,\quad 0\leq |\omega| < \pi,
\end{equation}
and shown in Fig.~\ref{fig:rdplots}. 
Moreover, let the distortion constraints be $D_C = 0.08$ and $D_S= 0.4$. 
%
Using Proposition~\ref{prop:stationary_RDF},  we have performed a simple grid search over $\lambda_1$ and $\lambda_2$. As $\lambda_1$ and $\lambda_2$ are varied, we compared the resulting side and central distortions given by~(\ref{spectra_constraints}) to the above mentioned distortion contraints $D_S$ and $D_C$. 
The noise spectra that resulted in distortions closest to the constraints are shown in Fig~\ref{fig:rdplots}. The spectra were obtained with $\lambda_1=0.2380$, and $\lambda_2 = 2.700$, which resulted in
$D_C = 0.0801$ and $D_S= 0.4000$. Moreover, when using these spectra
in~(\ref{theo:R}) the obtained per description rate is $R=0.7468$
bits/dim. In Fig.~\ref{fig:rdplots}, we have also shown the resulting side and central distortion spectra.
Note that for these values of $\lambda_1,\lambda_2$ and $S_X$, the set
of support frequencies are given by $|\omega|<
2.7173$. 

\section{Optimal Time-Domain Colored MD Coding}
\label{sec:mdscheme}

We now turn to the task of constructing a coding scheme. By Proposition~\ref{prop_Chen},  the test-channel of \figref{Ozarow_colored}, with optimum spectra $\Tp$ and $\Tm$, describes the optimal MD RDF $R(S_X,D_C,D_S)$. However, we would like a test channel that provides a recipe for a coding scheme, i.e., the additive noise elements should be replaced by quantizers. Unfortunately, the noises in \figref{Ozarow_colored} are colored and correlated with each other, making the task hard. Furthermore, we would like
to save the complexity of joint encoding of correlated samples by means of predictive
coding, as in DPCM.

In Section~\ref{sec:MD_test} we present an optimal test-channel with a single white additive noise, and in Section~\ref{sec:scheme} we present a corresponding coding scheme which ``replaces'' this noise with a quantizer. In order to prepare the ground for this test channel, we start with a simpler related problem, which is single-description coding under a spectral mask constraint on the reconstruction noise.

\subsection{Time-Domain Test Channel for a Distortion Mask}
\label{sec:mask}
We consider coding of a source subject to a maximum
distortion \emph{mask} $\{D(\ej)\}$. That is, we wish to compress a stationary
Gaussian source $X$ such that the spectrum of the reconstruction error
satisfies:\footnote{In principle, there is no guarantee that the
  reconstruction error is stationary. The condition can be thought of
  this way: pass the reconstruction error through a bank of filters,
  and measure the MSE as in the regular definition in
  Section~\ref{sec:prelim}; now take the limit of narrow band filters.} \beq{distortion_mask} S_{\hat X - X}(\ej) \leq D(\ej),\quad  -\pi < \omega \leq \pi . \eeq

Without loss
of generality\footnote{Otherwise, there is just wasted allowed
distortion which does not serve to reduce the rate.}, we assume that
$D(\ej)\leq \s, \forall \omega$. It is easy to verify, that the
minimum rate for this problem is given by 
\beq{RD_mask}
R\big(S_X,D\big) = \frac{1}{2}
\log\left(\frac{P_e(S_X)}{P_e\big(D\big)}\right). \eeq

\begin{figure*}[t]
\begin{center}
\psfrag{X}{$\scriptstyle X[n]$}
\psfrag{U}{$\scriptstyle U[n]$}
\psfrag{B}{$\scriptstyle B[n]$}
\psfrag{D}{$\scriptstyle D[n]$}
\psfrag{Y}{$\scriptstyle Y[n]$}
\psfrag{V}{$\scriptstyle V[n]$}
\psfrag{Z}{$\scriptstyle Z[n]$}
\psfrag{E}{$\scriptstyle E[n]$}
\psfrag{Et}{$\scriptstyle \tilde{E}[n]$}
\psfrag{Xh}{$\scriptstyle \hat{X}[n]$}
\psfrag{F(z)}{$\scriptstyle F(z)$}
\psfrag{A(z)}{$\scriptstyle \ A(z)$}
\psfrag{F*(z)}{$\scriptstyle \ F^{*}(z)$}
\psfrag{C(z)-1}{$\scriptstyle \ \ C(z)$}
\includegraphics{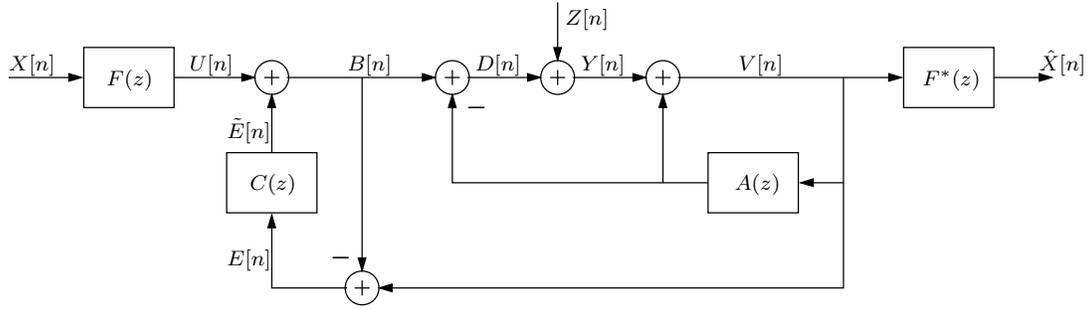}
\caption{A realization of the SD-RDF of a colored Gaussian source and subject to a distortion mask, by a test channel 
 channel consisting of pre/post filters, AWGN with noise shaping, and prediction.} \label{DPCM_DSM}
\end{center}
\end{figure*}

\figref{DPCM_DSM} presents a \emph{time-domain} test channel which
achieves this rate. Motivated by the ratio of entropy powers
\eqref{RD_mask}, we strive to achieve the optimal rate by the
combination of \emph{source prediction} in order to produce a prediction error of power $P_e(S_X)$, and \emph{noise
shaping} in order to shape the white quantization noise of power
$P_e(D)$ into the spectrum $D$.\footnote{An alternative
time-domain approach, is to accommodate for the distortion mask by
changing the pre and post-filters. However, we choose the
noise-shaping approach for the sake of extending this test channel to the
MD setting.}
These two tasks, we perform by a DPCM loop
\cite{zamir:2008} and a noise-shaping loop
\cite{ostergaard:2009}, respectively. In this test channel, $Z[n]$ is AWGN
of variance $P_e(D)$ and $A(z)$, which is given by~(\ref{eq:az}), is the optimal predictor
of the source spectrum $S_X$.\footnote{We assume that the optimal predictor $A(z)$ for the source spectrum exists. If not, then we may use the procedure outlined in Remark~\ref{rem:epsilon} in order to construct a predictor, which satisfies the assumption.} Moreover, the noise-shaping filter is given by: \begin{equation}\label{eq:C(z)}
C(z) = \frac{Q(z)}{1-Q(z)}.
\end{equation}
where $Q(z)$
is the optimal predictor for the distortion mask $\{D(\ej)\}$, see \eqref{spectral_decomposition}.
Note that
$E[n]$, the input to the noise-shaping filter, is equal to $Z[n]$.
The pre-filter $F(\ej)$ satisfies: \beq{pre_filter}|F(\ej)|^2 =
\frac{\s-D(\ej)}{\s}. \eeq

\begin{theorem} \label{thm_DPCM_DSM} For all source and distortion spectra satisfying $S_X(\ej)\geq D(\ej), \forall \omega$, and $\infty>P_e(S_X)\geq P_e(D)>0$, the channel of
\figref{DPCM_DSM} with the choices above, satisfies:
\beq{DPCM_DSM_dist} S_{\hat X - X}(\ej) = S_{V-U}(\ej) = D(\ej), \quad -\pi \leq \omega \leq \pi, \eeq
with the \emph{scalar} mutual information
$I(D[n];Y[n])=R\big(S_X,D\big)$ of \eqref{RD_mask}.
\end{theorem}

\begin{IEEEproof} Since $E[n]=Z[n]$, we have that 
\begin{align} \label{eq:n1} V[n] - U[n] &=  (1+c_n) * Z[n] \\ \label{eq:n2} Y[n] &= (1-a_n)*(U[n]+(1+c_n)*Z[n]). \end{align}
For the distortion spectra, the additive channel \eqref{eq:n1} means that $S_{V-U}(\ej) = D(\ej)$. By the choice of $F(\ej)$ and  $S_{\hat X - X}(\ej)$  follows as well.
For the mutual information, consider \eqref{eq:n2}: using the independence of $\{Z[n]\}$ from $\{U[n]\}$ and the choice of $F(\ej)$, $A(z)$ and $C(z)$, $Y[n]$ is white with variance $P_e(S_X)$. Moreover, $I(D[n];Y[n]) = h(Y[n]) - h(Y[n]|D[n]) =h(Y[n]) - h(Z[n])$, where $h(Z[n]) = P_e(D)$. 
\end{IEEEproof}

\begin{remark}
In the special case of a white distortion mask
$D$, the constraint becomes (by the water-filling principle)
equivalent to a regular quadratic distortion constraint. Indeed, the
channel collapses in this case to the pre/post filtered DPCM channel
of \cite{zamir:2008}. Much of the analysis there remains valid for
this problem as well. In particular, we can construct an optimal
coding scheme using this test channel, substituting the AWGN for an ECDQ,
and the scalar mutual information $I(D[n];Y[n])$ is also equal to
the directed mutual information $I(D[n]\rightarrow Y[n])$ and to the mutual information rate of the processes $\{X[n]\}$ and $\hat X[n]$.
\end{remark}

\subsection{Coloured MD Test Channel} \label{sec:MD_test}

Returning to the MD problem, it would be convenient to define the equivalent up-sampled noise
spectrum:
\beq{theta}
\tTheta(\ej) = \left\{
\begin{array}{ll}
            2\Theta_+\left(e^{j2\omega}\right), & \hbox{$|\omega| < \frac{\pi}{2}$}, \\
            2\Theta_-\left(e^{j2 \left(\omega-\frac{\pi}{2}\right)}\right), & \hbox{$\frac{\pi}{2} < \omega \leq \pi$}, \\
            2\Theta_-\left(e^{j2 \left(\omega+\frac{\pi}{2}\right)}\right), & \hbox{$-\pi \leq \omega < -\frac{\pi}{2}$},
        \end{array} \right.
        \eeq
where the lowpass and highpass spectra of $\tilde{\Theta}$ are
formed by $\{\Theta_+(e^{j2\omega})\}$ and $\{\Theta_{-}(e^{j2\omega})\}$,
which are compressed versions (by a factor of two) of the spectra
$\Theta_+ = \{\Tp\}_{\omega=-\pi}^{\pi}$ and $\Theta_-= \{\Tm\}_{\omega=-\pi}^{\pi}$,
respectively. These spectra can be taken to be the minimizers of \eqref{ParametricChenTianDiggaviRate}.
It is not hard to verify that the entropy power \eqref{eq:entropy_power}
of this spectrum is given by
\beq{theta_entropy_power}
P_e(\tTheta)
=
\exp \left(
\inthalf{ \log (2\sqrt{\Tp\Tm}) }
\right) .
\eeq
With this, the rate \eqref{ParametricChenTianDiggaviRate} can be expressed as
\begin{equation}
\label{ParametricChenTianDiggaviRate_tilde}
\frac{1}{2} \log \left( \frac{ P_e(S_X) }{ 2 \sqrt{P_e(\Theta_+)P_e(\Theta_-)}} \right) = \frac{1}{2} \log \left( \frac{ P_e(S_X) }{ P_e(\tTheta)} \right) .
\end{equation} Comparing to \eqref{RD_mask}, we see that the spectrum $\tTheta$ plays a similar role to that of the distortion mask $D$ in the previous section.

\begin{figure*}[t]
\begin{center}
\psfrag{X}{$\scriptstyle X[n]$}
\psfrag{U}{$\scriptstyle U[m]$}
\psfrag{B}{$\scriptstyle B[m]$}
\psfrag{D}{\hspace{-0.5mm}$\scriptstyle D[m]$}
\psfrag{Y}{\hspace{-0.5mm}$\scriptstyle Y[m]$}
\psfrag{V}{$\scriptstyle V[m]$}
\psfrag{Z}{$\scriptstyle Z[m]$}
\psfrag{E}{\hspace{-1mm}$\scriptstyle E[m]$}
\psfrag{Et}{\hspace{-1mm}$\scriptstyle \tilde{E}[m]$}
\psfrag{Xh}{$\scriptstyle \hat{X}_S[n]$}
\psfrag{Xc}{$\scriptstyle \hat{X}_C[n]$}
\psfrag{F(z)}{$\scriptstyle F(z)$}
\psfrag{A(z)}{$\scriptstyle A(z^2)$}
\psfrag{F*(z)}{$\scriptstyle \ F^{*}(z)$}
\psfrag{G*(z)}{$\scriptstyle \ G^{*}(z)$}
\psfrag{C(z)-1}{$\scriptstyle \ \ C(z)$}
\psfrag{LPF}{\scriptsize \ LPF}
\includegraphics[width=\textwidth]{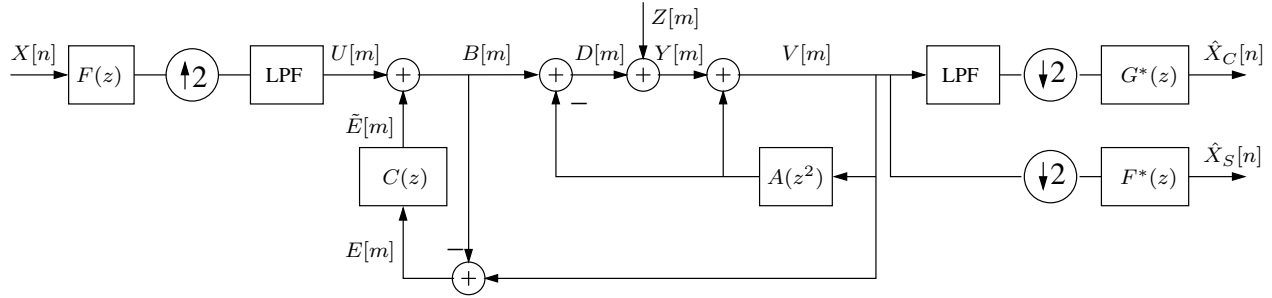}
\caption{A realization of the MD-RDF of a colored Gaussian source, by a test channel
consisting of pre/post filters, oversampling, AWGN with noise shaping, and
prediction. We use the index $n$
for sequences which are ``running'' at the source rate, and the
index $m$ when referring to the upsampled rate.}
\label{DPCM_DSM_2}
\end{center}
\end{figure*}

\figref{DPCM_DSM_2} shows the adaptation of the distortion-mask
equivalent channel to the MD problem. Following
\cite{ostergaard:2009}, we combine interpolation by a factor of two (upsampling and perfect low-pass filtering)
with the noise-shaping loop, forming a DSQ loop. 
$A(z)$ is the optimal predictor \eqref{eq:az} of
the spectrum $S_X(\ej)$.
It is worth pointing out that the predictor is applied
on the noisy output of the quantizer. Thus, the DPCM loop uses clean
predictions only in the limit of high resolutions,
see~\cite{zamir:2008} for details.
Note also that we apply an upsampled version of the source predictor, namely
$A(z^2)$.
The DSQ loop, on the other hand, works in the upsampled rate and the
noise-shaping filter $C(z)$ is given by~(\ref{eq:C(z)}) for the
spectrum $\tTheta(\ej)$ given by \eqref{theta}. Note that $E[m] = Z[m]$ in \figref{DPCM_DSM_2}, and we could equivalently
get the feedback noise $\tilde{E}[m]$ directly by passing the AWGN $Z[m]$ through the shaping filter $C(z)$.\footnote{The reason we drew the figure this
way is to make the connection to the coding scheme of \figref{high_res} more transparent.}
Recall from
\eqref{theta} that $\Theta_+(\ej)$ and $\Theta_-(\ej)$ now play
the roles of ``in-band noise'' and ``out of band noise''.
In the high-resolution case, the DSQ is independent of the shape of
the source
spectrum and uses a two-step spectrum as in
the white case, see \cite{ostergaard:2009} for details.
The additive Gaussian noise $Z[m]$ is white with variance \[P_e(\tTheta) = 2 \sqrt{P_e(\Theta_+)P_e(\Theta_-)}. \]
Finally, the filters $F(\ej)$ and $G(\ej)$ are chosen according to \eqref{filters}. The channel is completely characterized, then, by the choice of $\Tp$ and $\Tm$, which can be performed e.g. according to the optimization in Section~\ref{sec:spectral}. We show that regardless of the optimization, for any choice of spectra that satisfy \eqref{theta_region_spectra}, the rate $R$ per description is equal to the scalar mutual information
over the AWGN channel $D[m] \to Y[m]$ in Fig.~\ref{DPCM_DSM_2}, and explicitly given by (recall \eqref{explicit_minimization}):
\beq{before_minimization} 
\begin{split}
R(S_X,\Theta_+,\Theta_-)  &= \frac{1}{2} \log
\left(\frac{P_e(S_X)}{2\sqrt{P_e(\Theta_+)P_e(\Theta_-)}}\right) \\
&=
\inthalf{\frac{1}{2} \log \frac{S_X(\ej)}{2\sqrt{\Tp\Tm}}}. 
\end{split}
\eeq

\begin{remark} For a white
source, $A(z)=0$ and the channel reduces to the DSQ MD test channel of
\cite{ostergaard:2009}, while for optimal side distortion, $C(z)=0$,
and the channel reduces to an upsampled version of the DPCM
equivalent channel of \cite{zamir:2008}. \end{remark}

\begin{theorem} \label{thm_DPCM_DSM_2} The channel of
\figref{DPCM_DSM_2}, with the above choice of filters, satisfies:
\beqn{DPCM_DSM_2_dist} S_{\hat X_C - X}(\ej) & = & \Dc , \quad -\pi \leq \omega \leq \pi,
\nonumber \\ S_{\hat X_S - X}(\ej) & = & \Ds , \quad -\pi \leq \omega \leq \pi,
\eeqn 
where the distortion spectra were defined in
\eqref{spectra_constraints}, 
while the \emph{scalar} mutual
information $I(D[m];Y[m])$ equals the rate \eqref{before_minimization}.
\end{theorem}

\begin{IEEEproof}
The channel between $U[m]$ and $V[m]$ is identical to the one between $U[n]$ and $V[n]$ in \figref{DPCM_DSM}, with $A(z)$ replaced by $A(z^2)$. Thus, as in the proof of Theorem~\ref{thm_DPCM_DSM} we have:
\beq{eq:m1} V[m] - U[m] =  (1+c_m) * Z[m] \eeq  \beq{eq:m2} Y[m] = (1-\tilde a_m)*(U[m]+(1+c_m)*Z[m]), \eeq where $\tilde a_m$ corresponds to $A(z^2)$, i.e., equals $a_{m/2}$ for even $m$ and zero otherwise.

For the distortion spectra, the additive channel \eqref{eq:m1} means that $S_{V-U}(\ej) = \tTheta(\ej)$. By the choice of $|F(\ej)|^2$ and $|G(\ej)|^2$ and the equal phase condition (see \eqref{filters}),  and by the structure of rate conversion and low-pass filtering, it follows that 
\begin{align*} 
S_{\hat X_C - X}(\ej) &= \frac{\sx\Tp}{\sx - \Tm}
\nonumber \\ S_{\hat X_S - X}(\ej) &= \frac{1}{2} \left[
  \tTheta(e^{j2\omega}) + \tTheta(e^{j2\omega - j \pi})\right]\nonumber\\ 
&= \Tp + \Tm.
\end{align*}
The mutual information follows from \eqref{eq:m2}, similar to the proof of Theorem~\ref{thm_DPCM_DSM}.
\end{IEEEproof}

\subsection{Coding Scheme} \label{sec:scheme}

We now present a coding scheme based on the optimal time-domain test
channel. The scheme has the desirable property that the dependence
upon the specific source spectrum is limited to the choice of linear
operations (filters), while the quantization element itself is
generic. An alternative approach which achieves similar merits in the
frequency domain is suggested in \cite{chen:2009}. The choice between
time-domain and frequency-domain approaches is not specific to
multiple descriptions. For details about single-description quantization schemes based on DPCM and transform coding, see e.g. \cite{jayant:1984}. 

The underlying principle in transforming the test channel into a scheme is that an additive Gaussian white noise element in the test-channel may be replaced by an entropy-coded dithered quantizer (ECDQ), such that the rate of the quantizer is the scalar mutual information over the AWGN, and the variance of the (additive) quantization noise is the AWGN variance. This has been shown in~\cite{zamir:1996}, and extended to test channels with feedback loops in~\cite{zamir:2008}. We give a short account of these results, before turning to the MD scheme.

At all resolutions, an ECDQ can approach, at the limit of high lattice dimension $K\rightarrow\infty$,
a rate
\[ R = \frac{1}{2} \log \left(\frac{\sigma_Y^2}{\sigma_Z^2}\right), \] where $\sigma_Y^2$ and $\sigma_Z^2$ are the output and quantization noise variances, respectively. Furthermore, the quantization noise is independent of the quantizer input.\footnote{We will be using linear operations in our coding scheme, which are optimal in the high-dimensional limit, where the quantization noise becomes approximately Gaussian distributed (in a divergence sense) \cite{zamir:1996a}.}
Thus, in conjunction with optimal factors (i.e. Wiener estimation), the white Gaussian RDF is achieved \cite{zamir:1996}. In the presence of feedback loops, it is convenient to assume the
existence of a large number of identical and mutually
independent sources, equal to the lattice dimension $K$. These sources are treated
independently by the encoding/decoding scheme, except for the actual ECDQ which processes them
jointly. Thus we will only present the scheme for one source, but
the quantization noise has the properties of a high-dimensional ECDQ.
In a realistic setting, the multiple-source setting may be emulated by a single source which is divided into
$K$ long blocks and jointly encoded as $K$ parallel sources,
see~\cite{zamir:2008} for details.

\begin{figure*}[t]
\begin{center}
\psfrag{X}{$\scriptstyle X[n]$}
\psfrag{U}{$\scriptstyle U[m]$}
\psfrag{B}{$\scriptstyle B[m]$}
\psfrag{B1}{$\scriptstyle B_1[n]$}
\psfrag{B2}{$\scriptstyle B_2[n]$}
\psfrag{D1}{\hspace{-0.5mm}$\scriptstyle D_1[n]$}
\psfrag{D2}{\hspace{-0.5mm}$\scriptstyle D_2[n]$}
\psfrag{V}{$\scriptstyle V[m]$}
\psfrag{V1}{$\scriptstyle V_1[n]$}
\psfrag{V2}{$\scriptstyle V_2[n]$}
\psfrag{E}{\hspace{-1mm}$\scriptstyle E[m]$}
\psfrag{Q}{\hspace{-2mm}$\scriptstyle \mathcal{Q}(\cdot)$}
\psfrag{Y1}{\hspace{-0.5mm}$\scriptstyle Y_1[n]$}
\psfrag{Y2}{\hspace{-0.5mm}$\scriptstyle Y_2[n]$}
\psfrag{Et}{\hspace{-1mm}$\scriptstyle \tilde{E}[m]$}
\psfrag{Xh}{$\scriptstyle \hat{X}[n]$}
\psfrag{F(z)}{$\scriptstyle F(z)$}
\psfrag{A(z)}{$\scriptstyle \ A(z)$}
\psfrag{C(z)-1}{$\scriptstyle \ \ \ C(z)$}
\psfrag{LPF}{\scriptsize LPF}
\includegraphics{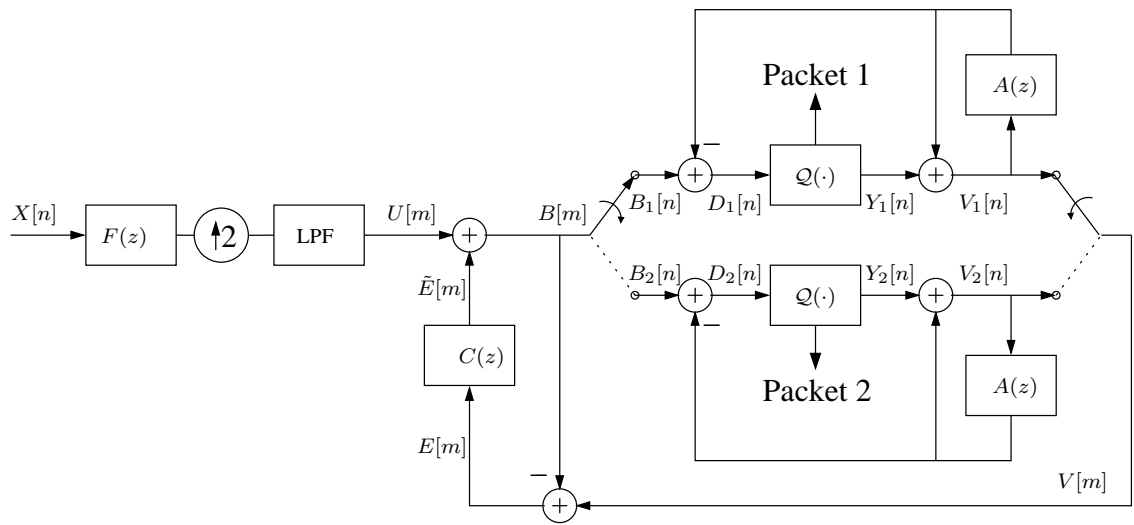}
\caption{Nested DSQ/DPCM MD encoder. The outer feedback loop $E[m]=V[m]-B[m]$ extracts the quantization noise,
that plays the role of the AWGN in the test channel of~\figref{DPCM_DSM_2}.}
\label{high_res}
\end{center}
\end{figure*}

\begin{figure*}[t]
\psfrag{X1}{$\scriptstyle \hat{X}_1[n]$}
\psfrag{X2}{$\scriptstyle \hat{X}_2[n-1]$}
\psfrag{Xc}{$\scriptstyle \hat{X}_C[n]$}
\psfrag{Y1}{$\scriptstyle Y_1[n]$}
\psfrag{Y2}{$\scriptstyle Y_2[n]$}
\psfrag{A(z)}{$\scriptstyle \ A(z)$}
\psfrag{LPF}{\scriptsize LPF}
\psfrag{F*(z)}{$\scriptstyle \ F^{*}(z)$}
\psfrag{G*(z)}{$\scriptstyle \ G^{*}(z)$}
\psfrag{z}{$\scriptstyle z^{-1}$}
\begin{center}
\includegraphics{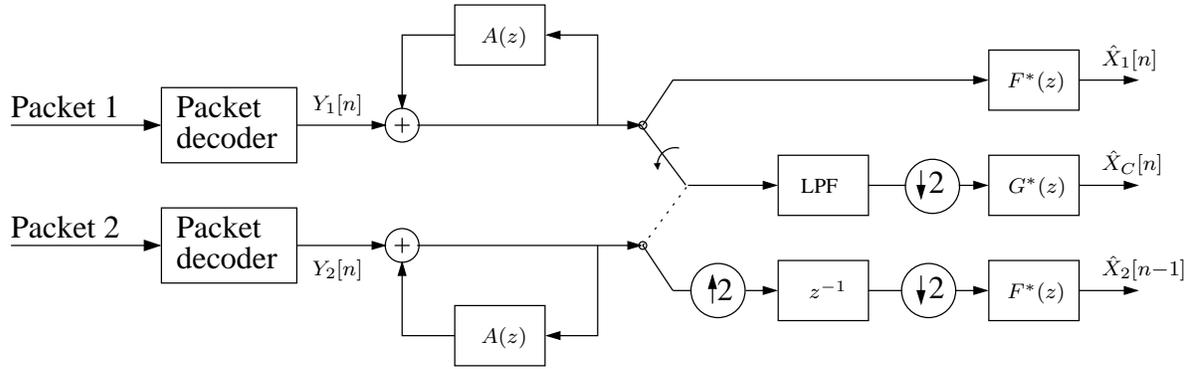}
\caption{DSQ/DPCM MD decoder.}
\label{high_res_decoder}
\end{center}
\end{figure*}

The encoder and decoder which materialize the optimal test channel
are presented in \figref{high_res} and \figref{high_res_decoder},
respectively. All of the switches in the encoder and the decoder are
synchronized.\footnote{It is to be understood that the switches change their positions with the upsampled rate ($m$).
Thus, in the encoder shown in Fig.~\ref{high_res}, the even samples $B_1[n]$ of $B[m]$ will go on the upper branch and the odd samples $B_2[n]$ will go on the lower branch.}
The up sampling operation followed by lowpass filtering introduces a half-sample delay on the odd samples.
This delay is corrected at the decoder by the delay operator $z^{-1}$ combined with the pair of up and downsamplers, see Fig.~\ref{high_res_decoder}.
The outputs of the quantizer blocks $\mathcal{Q}(\cdot)$ are the reconstructed values $Y_1[k]$ and $Y_2[k]$. Moreover, at each time $k$, the codeword of quantizer 1 (quantizer 2) is entropy-coded (conditioned upon the dither signal) and put into packet 1 (packet 2). The packet encoding operation is reversed at the decoder in order to obtain $Y_1[k]$ or $Y_2[k]$.

The operation of applying the predictor $A(z)$ to each description $V_1[n]$ and $V_2[n]$ in the encoder of Fig.~\ref{high_res}, is equivalent to applying the upsampled predictor $A(z^2)$ to the signal $V[m]$ in the test channel in Fig.~\ref{DPCM_DSM_2}. By using high-dimensional vector quantizers (ECDQs) with mutually independent dither sequences in the encoder,  the two resulting quantization noise sequences become statistically equivalent to the additive noise $Z[m]$ of the test channel. Consequently, the two descriptions $Y_1[n]$ and $Y_2[n]$ are
equivalent to the odd and even samples, respectively, of $Y[m]$ in
the equivalent test channel. Finally, the whole channel from the source
to the central and side reconstructions is equivalent to the channel
from $X[n]$ to $\hat X_C[n]$ and $\hat X_S[n]$, respectively.

Since we have seen that the mutual information in the test channel achieves (with optimized spectra $\Tp$ and $\Tm$) the colored MD RDF, the encoder/decoder pair of Figs.~\ref{high_res} and \ref{high_res_decoder} is able to achieve the complete RDF for stationary Gaussian sources at all resolutions and for any desired side-to-central distortion ratio.

\begin{remark}
In the scheme shown in Fig.~\ref{high_res}, the two prediction loops are embedded within a common noise shaping loop. Alternatively, one may alter the nesting order and let the common noise shaping loop be embedded within the two prediction loops. At high-resolution conditions, there is no loss of performance by switching the nesting order. However, at general resolution, the latter approach is suboptimal.
The reason is, that for white quantization noise, the DPCM loop also shows to the outside a total white noise
(by the basic DPCM equality~\cite{jayant:1984}), while the DSQ loop
shapes the noise. Since the DPCM loop assumes white noise for
optimality \cite{zamir:2008}, it cannot be built around the shaped DSQ noise.
\end{remark}


\section{Conclusions and Discussion}\label{sec:conclusion}
A parametric formulation of the two-description symmetric RDF for
stationary colored Gaussian sources and MSE was presented.
This result was established by providing a spectral domain
characterization of the optimum side and central distortion
spectra. For white Gaussian sources, the optimum spectrum of the distortion
is a two step function. For colored sources, the
optimum distortion spectrum is generally not piece-wise
flat but depends upon the source PSD and the desired resolution (i.e., the desired central and side distortion levels).
It was furthermore shown that the symmetric MD RDF could be achieved
by a time-domain approach based on oversampled prediction and noise-shaping.
The time domain implementation demonstrated that, at high resolutions,
it was possible to separate the mechanism responsible
for exploiting the source memory (DPCM) from the
mechanism controlling the MD coding parameters
(noise shaping).

The second part of the paper concerning the time-domain design, falls
under the theme of ``Shannon meets Wiener'' in QG coding
problems (cf. filtered ECDQ \cite{zamir:1996}, Wiener coefficients in lattice
decoding \cite{erez:2004}, DPCM for rate distortion \cite{zamir:2008},
MMSE estimation in Gaussian channels \cite{forney:2003}). The overall idea (of "Shannon meets Wiener) is to formulate
information theoretic solutions --- which are based on general random
codes and joint-typicality encoding/decoding --- as a combination of ``simple'' blocks:
filters, predictors, and white codebooks.

The paper also presented a new differential form of the symmetric version of Ozarow's test channel. While it is straightforward to extend this new differential channel to the asymmetric case, it is tedious (if possible) to use our approach to derive the optimal spectra for the asymmetric case. Moreover, it is not immediately clear that the time-domain implementation that we proposed will be optimal for the asymmetric case, since the resulting noise process in the outer feedback loop becomes cyclostationary  due to multiplexing two i.i.d. processes (dither signals) of different variances. Thus, the asymmetric case requires a different proof technique and a different time-domain strategy than what is required for the symmetric case.


\appendices


\section{Proof of Proposition~\ref{prop_white}}
\label{app:prop_white}
The fact that the rate expressions \eqref{rate_det2} and  \eqref{Ozarow_rate} are equivalent
follows easily by substituting the distortion expression \eqref{distortions} into \eqref{Ozarow_rate}.
We are thus left to show the equivalence of the triangular region \eqref{theta_region} with \eqref{non-degenerate}, under the transformation \eqref{distortions}. First, it is not hard to verify that the condition $0\leq D_C \leq D_S \leq \sigma_X^2$ (which is a weakened form of \eqref{non-degenerate}) together with \eqref{distortions} implies: \begin{subequations}\label{eq:simple_region}
 \begin{align} \Theta_- & \leq \sigma_X^2 \label{simple_region1} \\ \Theta_- & \geq 0 \label{simple_region2} \\  \Theta_+ & \geq 0 .  \label{simple_region3} \end{align}  \end{subequations}As the same follows from  \eqref{theta_region}, it is enough to check the equivalence under these conditions. Now, substituting \eqref{distortions} in \eqref{non-degenerate}, and noting that under \eqref{eq:simple_region} the transformation \eqref{distortions} is one-to-one, we see that \eqref{non-degenerate} is equivalent to:
\begin{subequations}
\label{proof_subequations}
\begin{align} 
\Theta_+ + \Theta_- & \leq \sigma_X^2 \label{proof1} \\
\frac{\sigma_X^2 \Theta_+}{\sigma_X^2 - \Theta_-} & \geq 2(\Theta_+ + \Theta_-) - \sigma_X^2 \label{proof2} \\
\frac{\sigma_X^2 - \Theta_-}{\sigma_X^2 \Theta_+} & \geq \frac{2}{\Theta_+ + \Theta_-} - \frac{1}{\sigma_X^2}.  \label{proof3} \end{align} \end{subequations}
By \eqref{eq:simple_region} all the denominators are non-negative. 
We now consider the following cases.
\begin{enumerate}
\item If \eqref{proof1} holds with equality, then all of the inequalities \eqref{proof_subequations} hold with equality. All of the points $\Theta_+ + \Theta_- = \sigma_X^2$ (with $\Theta_+$ and $\Theta_-$ non-negative due to \eqref{eq:simple_region}) are thus equivalent to $D_C=D_S=\sigma_X^2$. However, the rate \eqref{rate_det2} equals zero for $\Theta_+=\Theta_-=\sigma_X^2/2$ and is positive otherwise, thus it suffices to include this point in the minimization.
\item $\Theta_+=0$. By \eqref{simple_region2} and \eqref{proof2} we have $0\leq\Theta_-\leq \sigma_X^2/2$. It is equivalent to $D_C=0$, with infinite rate.
\item By \eqref{simple_region3} and \eqref{proof1}, the case $\sigma_X^2-\Theta_-=0$ is included in case 1. The cases $\Theta_+=0$ and $\Theta_++\Theta_-=0$ are included in case 2. Thus we may assume now that all denominators are positive. Rearranging \eqref{proof_subequations} by first multiplying the second and
  third inequalities by the terms $(\sigma_X^2-\Theta_-)$ and
 $\sigma_X^2 \Theta_+ (\sigma_X^2-\Theta_-)$, respectively, leads to:
\begin{subequations}
\begin{align} 
\sigma_X^2 - \Theta_+ - \Theta_- & \geq 0 \\
\left(\frac{\sigma_X^2}{2}-\Theta_-\right)(\sigma_X^2 - \Theta_+ -
\Theta_-) & \geq 0 \\ \label{eq:proof3a}
(\Theta_- - \Theta_+) (\sigma_X^2 - \Theta_+ - \Theta_-) & \geq 0.
 \end{align} 
 \end{subequations}
All three inequalities are satisfied if and only if $\sigma_X^2/2 \geq \tm
\geq \tp$ or $\sigma_X^2 - \tm - \tp = 0$. The former is the triangular region, while the latter is case 1 above. \end{enumerate} 

\hfill \IEEEQED


\section{Proof of Lemma~\ref{lem:sequence}}\label{app:lem:sequence}

Since all points are regular, the KKT conditions are necessary
conditions for optimality \cite{boyd:2004}. It follows that necessary conditions for optimality,
in addition to the constraints \eqref{constraints_start} --
\eqref{eq:noise_constraint}, are that all multipliers must be
non-negative, the complementary conditions must be satisfied:
\begin{align}
\mu_1(k)\Theta_+(k)  & = 0 , k=1,\dotsc, K \\
\mu_2(k)\left(\eta(k)/2 - \Theta_-(k) \right) & = 0 , k=1,\dotsc, K \\
\mu_3(k)\left(\Theta_-(k) - \Theta_+(k)\right) & = 0, k=1,\dotsc, K \\
\lambda_1 \left( \sum_{k=1}^K D_S\left(\eta(k),\tpk,\tmk\right) - D_S \right)&=0 \\
\lambda_2 \left( \sum_{k=1}^K D_C\left(\eta(k),\tpk,\tmk\right) - D_C \right)&=0,
\end{align}
and the variational conditions must be satisfied:
\begin{align*}
\frac{\partial \tilde J}{\partial \Theta_+(k)} = \frac{\partial \tilde J}{\partial \Theta_-(k)}=0,\quad k=1,\dotsc, K.
\end{align*}
The latter conditions can be written as:
\begin{align*}
&\frac{\partial \lk}{\partial \Theta_+(k)} \\
& = \frac{\partial \lk}{\partial \Theta_-(k)}=0,\quad k=1,\dotsc, K,
\end{align*} or explicitly as (recall \eqref{eq:diftheta+} and \eqref{eq:diftheta-}):
\begin{align}\label{eq:diftheta++}
-\frac{1}{4\tpk} + \lambda_1 + \lambda_2\frac{\eta(k)}{\eta(k)-\tmk} - \mu_{1}(k) + \mu_{3}(k) & = 0 \\
\label{eq:diftheta--}
-\frac{1}{4\tmk} + \lambda_1 + \lambda_2\frac{\eta(k)\tpk}{(\eta(k)-\tmk)^2} + \mu_{2}(k) - \mu_{3}(k) & = 0.
\end{align}

If $\lambda_1=0$ or $\lambda_2=0$, the problem reduces to SD
optimization, and it can be verified that the solution is nothing but
an alternative representation of the well-known SD reverse water-filling
solution. We concentrate therefore on the case $\lambda_1,\lambda_2>0$.

We start by noting that $\tpk=0$ means that (for any $\eta(k)>0$) the rate $R(\eta(k),\tpk,\tmk)$ is infinite. We may thus proceed by assuming w.l.o.g.\ that $\mu_1(k)=0, \forall k$, i.e., the corresponding constraint is never active. We proceed to check the other constraints.

\begin{enumerate}
\item $\mu_3(k)=0$, $\tmk=\eta(k)/2$, $0<\tpk<\eta(k)/2$. Substituting
  this case into \eqref{eq:diftheta++} and \eqref{eq:diftheta--} and imposing the non-negativity of the multiplier $\mu_{2}(k)$, we have:
\begin{align}
-\frac{1}{4\tpk} + \lambda_1 + 2 \lambda_2  & = 0 \label{equality_cond1} \\
-\frac{1}{2\eta(k)} + \lambda_1 + 4 \lambda_2\frac{\tpk}{\eta(k)}  & \leq 0. \label{ineq_cond1}
\end{align}
Substituting $\lambda_1$ from \eqref{equality_cond1} in \eqref{ineq_cond1} and rearranging, we have:
\[\left(\tpk-\frac{1}{8\lambda_2}\right)\left(\tpk-\frac{\eta(k)}{2}\right) \leq 0. \]
which yields $\tpk< \eta(k)/2$ only if \begin{align} \label{contra1} \lambda_2 > \frac{1}{4\eta(k)}. \end{align} On the other hand, substituting $\tpk$ from \eqref{equality_cond1} in \eqref{ineq_cond1} yields \begin{align} \label{contra2} 2\eta(\lambda_1+2\lambda_2) \leq 1. \end{align} For \eqref{contra1} and \eqref{contra2} to hold together it must be that $\lambda_1<0$, thus this option can never hold.

\item $\mu_2(k)=0$, $0<\tpk=\tmk<\eta(k)/2$. Substituting this case in the difference of \eqref{eq:diftheta++} and \eqref{eq:diftheta--} and imposing the non-negativity of the multiplier $\mu_{3}(k)$ yields $\tpk \geq \eta(k)/2$, which is in contradiction to the assumption.

\item $\tpk=\tmk=\eta(k)/2$. Substituting this case in the sum of
  \eqref{eq:diftheta++} and \eqref{eq:diftheta--} and imposing the
  non-negativity of the multiplier $\mu_{2}(k)$ leads to the condition
  $2\eta(k)(\lambda_1+2\lambda_2)\leq 1$.

\end{enumerate}

We have so far established that an optimal solution on the boundary
can only be: $\tpk=\tmk=\eta(k)/2$,
$2\eta(k)(\lambda_1+2\lambda_2)\leq 1$. In order to see that this is indeed a minimum, we check the Hessian. It is easy to see that
\begin{equation}
\frac{\partial^2}{\partial \tpk^2} \l = \frac{1}{4\tpk^2},
\end{equation}
\begin{equation}
 \frac{\partial^2}{\partial \tmk^2} \l = \frac{1}{4\tmk^2} + \frac{\lambda_2\eta(k)\tpk}{(\eta(k)-\tmk)^3},
\end{equation}
and
\begin{equation}
\frac{\partial^2}{\partial \tpk\partial \tmk} \l = \frac{\lambda_2\eta(k)}{(\eta(k)-\tm(k))^2}.
\end{equation}
To show that the determinant is positive it suffices to verify the following inequality:
\begin{align}\nonumber
\frac{1}{4\tpk^2}&\bigg(\frac{1}{4\tmk^2} +
\frac{\lambda_2\eta(k)\tpk}{(\eta(k)-\tmk)^3}\bigg) \\
& > \bigg(\frac{\lambda_2\eta(k)}{(\eta(k)-\tmk)^2}\bigg)^2,
\end{align}
which may be rewritten as
\begin{align} \nonumber 
&(4\tpk\tmk\lambda_2)^2  (\eta(k)-\tmk)(\lambda_2\eta(k)\tpk) \\ \label{eq:tmpo1}
&+ (\eta(k)-\tmk)^4 > (4\tpk\tmk\lambda_2\eta(k))^2.
\end{align}
The conditions  $2\eta(k)(\lambda_1+2\lambda_2)\leq 1$ and $\lambda_1>0$ imply that $\lambda_2\tmk < 1/8$. Moreover, since $\tpk\leq \tmk\leq \eta(k)/2$ we can upper bound the r.h.s.\ of \eqref{eq:tmpo1} by
\begin{equation}
 (4\tpk\tmk\lambda_2\eta(k))^2 \leq \frac{\eta(k)^4}{16}.
\end{equation}
Furthermore, since $\tmk\leq \eta(k)/2$ the second term on the l.h.s.\ of
\eqref{eq:tmpo1} is lower bounded by $\eta(k)^4/16$. Since the first term
on the l.h.s.\ is positive for $\lambda_2>0$, this proves that the
determinant is positive for $\lambda_1,\lambda_2>0$. Thus, we have
proven that at the optimal boundary solution
$\tmk=\tmk=\eta(k)/2$ we have $2(\lambda_1+2\lambda_2)\leq
1/\eta(k)$. We will now show the reverse direction, that for any
$\lambda_1>0, \lambda_2\geq 0$ such that $2(\lambda_1+2\lambda_2)\leq
1/\eta(k)$, the optimal solution must be this boundary solution. 
We first note that $2(\lambda_1+2\lambda_2)\leq
1/\eta(k)$ implies that $\lambda_1\leq 1/(2\eta(k)) - 2\lambda_2$,
which further implies that $\lambda_1 + \lambda_2 \leq
1/(4\eta(k))$. Then, it follows that 
\begin{align} \label{eq:tmjleq}
\tmk &\geq \tpk \\  \label{eq:opt_sol_tmp}
&= \frac{\eta(k) - \tmk}{ 4(\lambda_1 + \lambda_2)\eta(k) -
  4\lambda_1\tmk} \\ \label{eq:sum_lambda}
&\geq  \frac{\eta(k) - \tmk}{1 -
  4\lambda_1\tmk} \\ \label{eq:nonnegative_lambda}
&\geq \eta(k) - \tmk,
\end{align}
where \eqref{eq:opt_sol_tmp} follows from \eqref{eq:theta1}, \eqref{eq:sum_lambda} is due to using
$\lambda_1+\lambda_2\leq 1/(4\eta(k))$ and
\eqref{eq:nonnegative_lambda} follows since $0\leq 4\lambda_1\tmk \leq
2\tmk/\eta(k) \leq 1$. However, comparing \eqref{eq:tmjleq} and
\eqref{eq:nonnegative_lambda} shows that we require $\tmk \geq
\eta(k)/2$ in addition to the constraint $\tmk\leq \eta(k)/2$, which is only satisfied at
$\tmk=\eta(k)/2$. As shown above, $\tmk=\eta(k)/2$ implies $\tpk=\eta(2)/2$, which proves the claim. Interestingly, inserting $\tpk=\tmk=\eta(k)/2$ into \eqref{eq:opt_sol_tmp} yields $\tpk=\frac{1}{4(\lambda_1+2\lambda_2)} = \eta(k)/2$, or equivalently, $2(\lambda_1+2\lambda_2) = 1/\eta(k)$. 
  
Let us now consider the second part of the lemma, i.e., if $2\eta(k)(\lambda_1+2\lambda_2) >1$, then the optimal
pair of noise variances $\tpk, \tmk$ is strictly within the triangular
support region. To prove this, we will show that in this case, the
Lagrangian is
reduced by moving from the boundary and into the interior. Thus, there
must be at least one minimum within the interior. 

Recall that the boundary of the triangular optimization region \eqref{theta_region_spectra} (see also Fig.~\ref{fig:boundary}) is given by the union of the following three faces:
\begin{enumerate}
\item $\tpk =0, 0\leq \tmk\leq \eta(k)/2$.
\item $0<\tpk = \tmk \leq \eta(k)/2$.
\item $\tmk = \eta(k)/2, 0<\tpk\leq\tmk$.
\end{enumerate}
We now consider the behavior of the Lagrangian $\lk$ near these faces.
\begin{enumerate}
\item If $\tpk = 0$, then $R(k)\to \infty$ for any non-negative
$\lambda_1,\lambda_2$ and $\tmk<\eta(k)$. Thus, moving from the boundary
to the interior can only reduce the Lagrangian.

\item Let $0<\tpk=\tmk<\eta(k)/2$.
The (unnormalized) directional derivative, i.e., the ``normal'' vector
to the surface $\tpk = \tmk$, which
points into the support region (towards the interior), is given by the
following difference of partial derivatives: 
\begin{equation} \label{eq:directional}
\begin{split}
&\frac{\partial \lk }{\partial \tmk} \\
& \quad - \frac{\partial \lk }{\partial \tpk} \\
&= \frac{\lambda_2 \eta(k)}{\eta(k)-\tpk}\bigg( \frac{2\tpk-\eta(k)}{\eta(k)-\tpk}  \bigg),
\end{split}
\end{equation}
where we have used~(\ref{eq:diftheta+}) and~(\ref{eq:diftheta-}), substituting $\tpk= \tmk$.
Since $\tpk <\eta(k)/2$ it follows that~(\ref{eq:directional}) is always
negative (for $\lambda_2>0$). Thus, if $\lambda_2>0$ 
then $\lk$ can be reduced by going away from the boundary
and into the interior, i.e., by letting $0<\tpk<
\tmk<\eta(k)/2$. On the other hand, if $\lambda_2=0$,
then  \eqref{eq:directional} is
zero, $D_S(k) = 2\tpk$ and  the rate is $R(k)=\frac{1}{4\pi}\log\left(\frac{\eta(k)}{D_S(k)}\right)$,
which is optimal for the side reconstructions, i.e., each side
reconstruction is on the SD RDF.
Finally, if $\lambda_2>0$ but $\tpk=\tmk=\eta(k)/2$, then the distortions satisfy $D_s(k) = D_c(k)=\eta(k)$
and the rate is $R(k) = 0$, which is a trivial solution.

\item For $\tmk=\eta(k)/2$, 
the directional derivative is given by
\begin{equation}\label{eq:directional1}
\begin{split}
&\left.\frac{\partial \lk}{\partial \tmk)}\right|_{\tmk=\eta(k)/2} \\
&=
-\frac{1}{2\eta(k)} + \lambda_1 +\frac{4\lambda_2\tpk}{\eta(k)},
\end{split}
\end{equation}
which is non-negative for $2\eta(k)(\lambda_1+2\lambda_2) >1$. 
To see this, 
we form the inequality
\begin{equation}\label{eq:directional2}
\frac{\partial \lk}{\partial  \tmk}\big|_{\tmk=\eta(k)/2} \geq 0.
\end{equation}
From \eqref{eq:theta1} it follows that $\tpk  =
1 / 4(\lambda_1 + 2\lambda_2)$, which when used in
\eqref{eq:directional2} leads to
\begin{equation}\label{eq:directional3}
2\eta(k)\lambda_1(\lambda_1 + 2\lambda_2)   \geq \lambda_1.
\end{equation}
We now use that $\tmk=\eta(k)/2$ and the condition $2\lambda_1 + 4\lambda_2 > 1/\eta(k)$, which when inserted in
\eqref{eq:directional3} confirms that \eqref{eq:directional1} is
positive for $\lambda_1>0$ and zero for $\lambda_1=0$. 
Let $\lambda_1 = 0, \lambda_2>0$ and $\tmk=\eta(k)/2$. Then $\tpk = 1/(8\lambda_2), D_C(k) = 1/(4\lambda_2)$, and $D_S(k) = \eta(k)/2 + D_C(k)/2$. Moreover, the per description rate is $R(k) = \frac{1}{4}\log_2(\eta(k)/D_C(k))$. As expected, this is an optimal operating regime for single-description coding at rate $2R(k)$ and distortion $D_C(k)$. 
\end{enumerate}

We have thus proved that in the interior of the support region, $\lk$ is
differentiable. Moreover, when $2\eta(k)(\lambda_1+2\lambda_2) >1$ we
have shown that $\lk$
decreases for $\lambda_1,\lambda_2>0$ when moving into the interior
from any boundary point. Thus, the Lagrangian must indeed have
at least one minimum within the interior. This completes the proof of
the lemma. \hfill \IEEEQED


\section{Proof of Lemma~\ref{lem:unique_sol}}\label{app:lem:unique_sol}

We know from Lemma~\ref{lem:sequence} that since we only consider
supported levels (see Def.~\ref{def:support}), the  global minimum is within interior of the support
region~\eqref{theta_region_spectra}. Moreover, we know that the
minimum must be a stationary solution given by a root of \eqref{eq:psipoly1}, since the Lagrangian is
differentiable in the support region. Depending on the sign of $\Xi$,
\eqref{eq:psipoly1} may have more than one root. 
Thus, to prove the lemma we have
to show that only one of the roots of \eqref{eq:psipoly1}  is an
optimal solution and identify that root. To do so, we will show that whenever
\eqref{eq:psipoly1} has more than one root, only one of them 
will be in the support region and this must be then be optimal solution. 

Let $a_0,\ldots,a_3$ denote the coefficients of
the third-order polynomial in \eqref{eq:psipoly1}, and recall the
definitions of $\Xi$, $p$, $q$, and $\phi$ given by \eqref{eq:Xi} --
\eqref{eq:phi}, respectively. Let $s_1 = \sqrr{q + \sqr{\Xi}}$
and $s_2 = \sqrr{q - \sqr{\Xi}}$. Then, the three roots of the polynomial are given by (see e.g.,~\cite{abramovitz:1973}):
\begin{align}\label{eq:x1}
x_1 &= (s_1+s_2) -\frac{a_2}{3} \\ \label{eq:x2}
x_2 &= -\frac{1}{2}(s_1+s_2) -\frac{a_2}{3} + \frac{i\sqr{3}}{2}(s_1-s_2) \\ \label{eq:x3}
x_3 &= -\frac{1}{2}(s_1+s_2) -\frac{a_2}{3} - \frac{i\sqr{3}}{2}(s_1-s_2).
\end{align}
If $\Xi>0$, then there is one real root and two complex roots.
If $\Xi<0$, there are three real distinct roots. Finally, if $\Xi=0$, there is a single real triple root (if $q=0$) or one real root and one real double root (if $q\neq 0$)~\cite{abramovitz:1973}.
Thus, for every choice of $(\lambda_1,\lambda_2)$, one may identify
the admissible solutions
of~(\ref{eq:x1}) -- (\ref{eq:x3}), i.e., the ones that are inside the triangular region~(\ref{theta_region_spectra}) (recall Figure~\ref{fig:boundary}).

It easy to show that the discriminant \eqref{eq:Xi} satisfies:
\begin{equation} \label{eq:Xi-xi}
\Xi=-\frac{\eta^{4}(4\lambda_1^2\eta^{2}+1)}{432\lambda_1^6}(\lambda_2-\xi^\Xi_0)(\lambda_2-\xi^\Xi_1)(\lambda_2-\xi^\Xi_2)(\lambda_2-\xi^\Xi_3),
\end{equation}
where $\{\xi^\Xi_i\}$ are the four real roots of $\Xi$ given by $\xi^\Xi_0=0, \xi^\Xi_1=-\lambda_1$,
\begin{align}\label{eq:xi2}
\xi^\Xi_2 &= -\frac{2\eta\lambda_1 + 8\eta^{3}\lambda_1^3 - 16\eta^{2}\lambda_1^2 -3 +2\sqr{2(2\eta^{2}\lambda_1^2 + 1)^3}}{4\eta(4\eta^{2}\lambda_1^2+1)}, \\ \label{eq:xi3}
\xi^\Xi_3 &= -\frac{2\eta\lambda_1 + 8\eta^{3}\lambda_1^3 - 16\eta^{2}\lambda_1^2 -3 -2\sqr{2(2\eta^{2}\lambda_1^2 + 1)^3}}{4\eta(4\eta^{2}\lambda_1^2+1)}.
\end{align}

Since we only have to consider non-negative multipliers, $\xi_1^\Xi\leq 0$.
Clearly, $\xi^\Xi_2<\xi^\Xi_3$. The following lemma states the signs of $\xi_2^\Xi$ and $\xi_3^\Xi$.

\begin{lemma}\label{lem:sign-xi3}
For $\lambda_1>0, \xi^{\Xi}_3 >0$. Moreover, the sign of $\xi_2^\Xi$~(\ref{eq:xi2}) is given by:
\begin{equation}\label{eq:signofD}
sign(\xi^\Xi_2)= sign\left( \frac{1}{4\eta} - \lambda_1 \right).
\end{equation}
\end{lemma}
\begin{IEEEproof}
We first show that $\xi_3^\Xi$~(\ref{eq:xi3}) is non-negative. To do so,
we show that
\begin{equation}
2\eta\lambda_1 + 8\eta^{3}\lambda_1^3 - 16\eta^{2}\lambda_1^2 -3 -2\sqr{2(2\eta^{2}\lambda_1^2 + 1)^3} < 0,
\end{equation}
which means that~(\ref{eq:xi3}) is positive.
Let $\varphi_1= 2\eta\lambda_1 + 8\eta^{3}\lambda_1^3$ and \\ $\varphi_2 =
2\sqr{2(2\eta^{2}\lambda_1^2 + 1)^3}$ and notice that it is enough to show that $\varphi_1<\varphi_2, \forall \omega$.
Since $\varphi_1$ and $\varphi_2$ are both positive functions, we may work on their squares, i.e., $\varphi_1^2= 4\eta^{2}\lambda_1^2 + 32\eta^{4}\lambda_1^4 + 64\eta^{6}\lambda_1^6$ and
$\varphi_2^2 = 64\eta^{6}\lambda_1^6 + 192\eta^{4}\lambda_1^4 + 192\eta^{2}\lambda_1^2 + 64$. Forming the inequality $\varphi_2^2>\varphi_1^2$ and collecting similar terms yields $64>-188\eta^{2}\lambda_1^2 -160\eta^{4}\lambda_1^4$ which is always satisfied for $\lambda_1\in\mathbb{R}$. This proves the first part of the lemma.

We now consider the sign of $\xi_2^\Xi$~(\ref{eq:xi2}).
Let $\varphi_1=2x + 8x^3 - 16x^2 -3$ and $\varphi_2=2\sqr{2(2x^2 + 1)^3}$.
The discriminant of $\varphi_1$ is strictly positive so $\varphi_1$ has only a single real root, which is located at $\xi=1.96973$ where we note that $\xi>\frac{1}{4}$. Moreover, $x=0\Rightarrow \varphi_1=-3$ and it follows that $\varphi_1<0$ for $x<\xi$ and $\varphi_1>0$ for $x>\xi$. Notice also that $\varphi_2>0$ for $x>0$.

At this point we let $h=\varphi_1^2 - \varphi_2^2 = -256 (z-\frac{i}2)(z+\frac{i}2)(z-\frac{1}4)^3$,
which is a fifth-order polynomial having a pair of complex conjugate roots at $x=\pm i/2$ and a real (triple) root at $x=1/4$. Thus, $h$ crosses the real line only once. Since $h=1$ for $x=0$ it follows that $h>0$ for $x<1/4$ and $h<0$ for $x>1/4$. Furthermore, $h=0$ for $x=1/4$.

Since $h>0$ for $x<1/4$ it follows that $\varphi_1^2>\varphi_2^2$ which implies that $\varphi_1+\varphi_2<0$ since $\varphi_1<0$ for $x<1/4$. The first case of~(\ref{eq:signofD}) now follows by inserting $x=\lambda_1\eta$ in $\varphi_1$ and remembering the additional sign from~(\ref{eq:xi3}). Since $h=0$ implies that $\varphi_1+\varphi_2=0$, it immediately follows that $sign(\xi^\Xi_2)=0$ for $x=\lambda_1\eta=1/4$. Finally, for $h<0$ we have $\varphi_2^2>\varphi_1^2$ which implies that $\varphi_2>\varphi_1$ since $\varphi_2$ is positive and it follows that $\xi^\Xi_2<0$ for $\lambda_1\eta>1/4$. This proves the remaining parts of the lemma.
\end{IEEEproof}

\begin{figure}

    \begin{center}
      \psfrag{D}{$\Xi$}
      \psfrag{l2}{$\lambda_2$}
      \psfrag{x0}{$\xi^\Xi_0$}
      \psfrag{x1}{$\xi^\Xi_1$}
      \psfrag{x2}{$\xi^\Xi_2$}
      \psfrag{x3}{$\xi^\Xi_3$}
      \psfrag{x12}{$\xi^\Xi_{1,2}$}
      \psfrag{x21}{$\xi^\Xi_{2,1}$}
\subfigure[$\lambda_1<\frac{1}{4\eta}$]{\includegraphics[width=6cm]{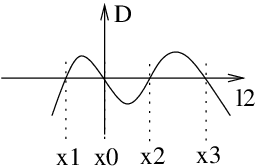}}
  \subfigure[$\lambda_1>\frac{1}{4\eta}$]{\includegraphics[width=6cm]{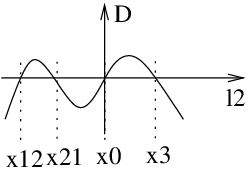}}
  \caption{The possible zero locations for the discriminant $\Xi$ as a function of $\lambda_2$.}
  \label{fig:roots_of_Xi}
  \end{center}
\end{figure}

In Fig.~\ref{fig:roots_of_Xi} we illustrate the possible sign behavior of $\Xi$ as a function of $\lambda_2$, using Lemma~\ref{lem:sign-xi3} and the fact that
by \eqref{eq:Xi-xi}, $\lim_{\lambda_2\rightarrow \pm\infty} \Xi = -\infty$. Building on this, we prove Lemma~\ref{lem:unique_sol} by considering the following three cases.

\subsection{Negative Discriminant}\label{sec:negative_disc}
In this case
$\Xi = q^2 + p^3 <0$ and we have three real solutions. It is easy to see that we must have $p<0$ and $|p|^3>q^2$. Let $z_1=q + \sqr{\Xi} = q + i\sqr{-\Xi}$
and $z_2=q - \sqr{\Xi} = q - i\sqr{-\Xi}$ and notice that $s_i = \sqrr{z_i}, i=1,2.$ Since $\Xi, p<0$, it follows that $|z_1| = |z_2| = \sqr{-p^3} = \sqr{|p|^3}$. 
Moreover, the phase $\phi_i$ of $z_i$ is given by $\phi_1=\arctan(\sqrt{-\Xi},q)$ and $\phi_2=-\phi_1$, respectively, since $\arctan$ is symmetric. Note that $\phi_1 \in [0;\pi]$ since $\sqr{-\Xi}> 0$ . 

With this, it is easy to show that the solutions (roots), $\{x_i\}_{i=1}^3,$ as given by~(\ref{eq:x1})--(\ref{eq:x3}), can be written as
\begin{align}\label{eq:x1t}
x_1 &= 2\sqr{|p|}\cos(\phi_1/3) -\frac{a_2}{3}, \\ \label{eq:x2t}
x_2 &= -\sqr{|p|}\big( \cos(\phi_1/3) +\sqr{3}\sin(\phi_1/3) \big) -\frac{a_2}{3}, \\ \label{eq:x3t}
x_3 &= -\sqr{|p|}\big( \cos(\phi_1/3) -\sqr{3}\sin(\phi_1/3) \big) -\frac{a_2}{3}.
\end{align}

We note that
\begin{align}\label{eq:zeta1}
\min_{\zeta \in [0;\pi]} 2\cos(\zeta/3) \geq  \max_{\zeta \in [0;\pi]}  -(\cos(\zeta/3)-\sqr{3}\sin(\zeta/3))
\end{align}
and that
\begin{align}\nonumber
&\min_{\zeta \in [0;\pi]}   -(\cos(\zeta/3)-\sqr{3}\sin(\zeta/3)) \\  \label{eq:zeta2}
 \geq & \max_{\zeta \in [0;\pi]}  -(\cos(\zeta/3)+\sqr{3}\sin(\zeta/3)),
\end{align}
which implies that $x_1\geq x_3\geq x_2$  for any pair $(\lambda_1,\lambda_2)$. In \eqref{eq:zeta1}, the minimum of the lhs.\ is clearly $1$. On the other hand, $-\cos(\zeta/3)$ is negative, increasing in $\zeta$, and having a maximum of $-1/2$. Moreover, $\sqrt{3}\sin(\zeta/3)$ is positive, increasing, and with a maximum of $3/2$. Thus, the rhs.\ is upper bounded by $-1/2 + 3/2$ and the lhs.\ is lower bounded by $1$. In \eqref{eq:zeta2}, the minimum of the lhs.\ is clearly $-1$, since both $-\cos(\zeta/3)$ as well as $\sqrt{3}\sin(\zeta/3)$ have their minima at $\zeta=0$. On the other hand, it is easy to show that $-\frac{\partial}{\partial \zeta} \sqrt{3}\sin(\zeta/3) \geq \frac{\partial}{\partial \zeta}\cos(\zeta/3)$, which means that the rhs.\ is a decreasing function in $\zeta$. Thus, inserting $\zeta=0$ yields the maximum, which is $-1$. This shows that the minimum of the lhs.\ is greater than or equal to the maximum of the rhs.

Let us now consider the solution $x_3$ given by~(\ref{eq:x3t}). From Lemma~\ref{lem:x3_large} below, it follows that $x_3>\eta/2$, which violates the spectral constraint. Moreover, since $x_1>x_3$, we deduce that $x_2$
given by~(\ref{eq:x2t}) is the only admissible candidate solution.

\begin{lemma}\label{lem:x3_large}
Let $\Xi<0$. Then, for any positive $\lambda_1$ and $\lambda_2, x_3>\eta/2$, where $x_3$ is given by~(\ref{eq:x3t}).
\end{lemma}

\begin{IEEEproof}
Let us first assume that $q<0$, which implies that $\phi_1=\arctan(\sqrt{-\Xi}/q) \in [\pi/2, \pi]$. However, for $\pi\geq \phi_1 \geq \pi/2, \cos(\phi_1/3)-\sqrt{3}\sin(\phi_1/2) \leq 0$. Thus, $-\sqrt{|p|}(\cos(\phi_1/3)-\sqrt{3}\sin(\phi_1/3)) \geq 0$. Since $a_2/3 > 2\eta/3$, it is clear that $x_3>\eta/2$ as was to be shown.  Now let $q=0$, in which case $\phi_1=\pi/2, (\cos(\phi_1/3)-\sqrt{3}\sin(\phi_1/3))  = 0$, which further implies that $x_3>\eta/2$. We may therefore proceed assuming $q> 0$. 

For $\lambda_1\geq 1/\eta$ and $\lambda_2\geq 1/\eta$, we will show that $\frac{\partial^2}{\partial \lambda_2^2} \phi_1 >0$, which implies that $\phi_1$ is convex in $\lambda_2$. 
In particular, $\frac{\partial^2}{\partial \lambda_2^2} \phi_1 = p_1\varphi_1$, where $p_1$ is some everywhere positive function of $\lambda_2$ and $\varphi_1$ is a $7th$-order polynomial in $\lambda_2$. 
To show that $\frac{\partial^2}{\partial \lambda_2^2} \phi_1>0$, it is therefore enough to show that $\varphi_1>0$. It is straightforward to check that $\frac{\partial^{7}}{\partial \lambda_2^{7}}\varphi_1 > 0$ for all $\lambda_2>0$ and $\lambda_1\geq  1/\eta$. Thus, $\frac{\partial^{6}}{\partial \lambda_2^{6}}\varphi_1$ is an increasing function in $\lambda_2$. Since we are only interested in the interval $\lambda_2\geq 1/\eta$, it follows that $\frac{\partial^{6}}{\partial \lambda_2^{6}} \varphi_1 |_{\lambda_2=1/\eta}$ yields the minimum, which is positive. It therefore also follows that $\frac{\partial^{5}}{\partial \lambda_2^{5}} \varphi_1$ is an increasing function. Evaluating $\frac{\partial^{5}}{\partial \lambda_2^{5}} \varphi_1$ at $\lambda_2=1/\eta$ yields a positive value. This shows that $\frac{\partial^{4}}{\partial \lambda_2^{4}} \varphi_1$ is increasing. 
We repeat this procedure for $\frac{\partial^{3-i}}{\partial \lambda_2^{3-i}}\varphi_1$ and for $i=0,\dotsc, 2$,  to show that they are all increasing and positive functions, which implies that $\varphi_1$ is an increasing function. Finally, evaluating $\varphi_1$ at $\lambda_2=1/\eta$ gives a positive value, which proves that $\varphi_1$ is positive in the interval $\lambda_2\geq 1/\eta$, and we conclude that $\phi_1$ is convex in this interval.

Assume $g$ is some affine function in $\lambda_2$ and define $f\triangleq\cos(\phi/3)-\sqrt{3}\sin(\phi/3)$.
Then, by comparing terms of $\frac{\partial^2}{\partial \lambda_2^2} gf$ and recalling that $\phi_1$ is convex and that we only need to consider $\phi_1 \in [0,\pi/2]$, it is easy to show that $\frac{\partial^2}{\partial \lambda_2^2} gf<0$, which implies concavity.  Note also that for $\lambda_1,\lambda_2 \geq 1/\eta$, $\frac{\partial^2}{\partial \lambda_2^2} \sqr{-p} = \frac{c\,(4\eta^{2}\lambda_1^2-1)}{\sqr{-p^3}}$, for some everywhere positive function $c$ (which is independent of $\lambda_2$). Thus, $\sqrt{-p}$ is convex in $\lambda_2$. Since the slope is increasing in $\lambda_2$, we can take the limit $\lambda_2\to\infty$ to show that the maximum slope of $\sqr{|p|}$ is $\frac{\eta}{3\lambda_1}$. It follows that we can upper bound $\sqrt{|p|}$ by the affine function $g\triangleq c_0 + \frac{\eta}{3\lambda_1}\lambda_2$, where $c_0=\frac{1}{12\lambda_1}(4\eta\lambda_1-1)$. 

We are now in a position to lower bound $x_3$, i.e.,  $x_3 > -fg - a_2/3$, where both $-fg$ and $-a_2/3$ are convex in $\lambda_1,\lambda_2\geq 1/\eta$. 
Since $\lim_{\lambda_2\to \infty}[\frac{\partial}{\partial \lambda_2}( -fg - a_2/3)] = 0$, the limit $\lambda_2\to\infty$ of $-fg -a_2/3$ must be the minimizer, which is given by
\begin{equation}
 \lim_{\lambda_2\to\infty} -fg - a_2/3 = \frac{1}{6\lambda_1} + \frac{\eta}{3} + \frac{1}{4\lambda_1}\sqrt{4\eta^2\lambda_1^2+1}  > \eta/2.
\end{equation}

Let us now consider the case $\lambda_2<1/\eta$ and $\lambda_1\geq 1/\eta$. For $\lambda_1\geq 1/(4\eta)$, we must have $\lambda_2>\xi_3^\Xi$ in order to have $\Xi<0$. However, $\xi_3^\Xi\geq 1/\eta$. Thus, we only need to check the final case of $\lambda_1<1/(4\eta)$ in which case we must have $\lambda_2<\xi_2^\Xi$. By comparing terms it may readily be verified that $\xi_2^\Xi<1/(4\eta)-\lambda_1/2$. Moreover, in this interval $\sqrt{-p}$ is a decreasing positive concave function (whenever $p$ is negative as required for $\Xi<0$). Thus, we may upper bound $\sqrt{-p}$ in this interval by $\sqrt{-p}|_{\lambda_2=0}$. With this we can lower bound $x_3$ by  $-\sqrt{-p}|_{\lambda_2=0} - a_2/3 = \eta/3 + 1/(6\lambda_1) > \eta/2$.  
This proves the lemma.

\end{IEEEproof}

\subsection{Positive Discriminant}
In this case $\Xi>0$ and we have only a single real solution given by
$x_1$ (\ref{eq:x1}).

\subsection{Zero Discriminant}\label{sec:zero_disc}

In this case $\Xi=0$, which is possible if $-p^3= q^2$. Thus, either
$p=q=0$ or $p<0$. If $q=0$,
there is a single real triple root ($x_1$). We therefore only need to consider
the case where $q\neq 0$ in which case there is a single real root
$x_1$ and a real double root $x_2=x_3$, since $s_1=s_2$. 
The zeros of $\Xi$ for which $\lambda_2>0$ are located at $\lambda_2=\xi_2^\Xi,\xi_3^\Xi$, where
the former is positive only if $\lambda_1<\frac{1}{4\eta}$. By inserting $\lambda_2=\xi_3^{\Xi}$ into $q$ it easy to verify that $q<0$. On the other hand, $q|_{\lambda_2=\xi_2^{\Xi}}$ has only one positive zero at $\lambda_1 = 1/(4\eta)$, and it is easy to check that $q_{\lambda_2=\xi_2^{\Xi}}<0$ for $\lambda_1<1/(4\eta)$. Thus, we may proceed assuming $q<0$. However, then clearly $s_i<0,
i=1,2$, which implies that $x_2=x_3>\eta/2$ since $-a_2/3
>\eta/2$. The only admissible solution is therefore $x_1$.

\section{Proof of Proposition~\ref{prop:hr}.}\label{app:proof:prop:hr}
\subsection{Case $\lambda_1>0$ and $\lambda_2\gg 1$}\label{sec:case1}
In this case, we note that
\begin{equation}\label{eq:Xi_order}
\Xi(\ej) = -\frac{\lambda_2^4\sxx{4}}{432\lambda_1^6}(1 + 4\sxx{2}\lambda_1^2) + \mathcal{O}\big( \frac{\lambda_2^3}{\lambda_1^{3}}\big).
\end{equation}
It follows that $\Xi(\ej)<0$ for large $\lambda_2$. Furthermore,
\begin{equation}
q(\ej) = \frac{\lambda_2^3 \sxx{3}}{27\lambda_1^3} + \mathcal{O}\big( \frac{\lambda_2^{2}}{\lambda_1^2} \big)
\end{equation}
and $\phi_1(\ej) = \arctan\left(\frac{\sqr{-\Xi(\ej)}}{q(\ej)}\right).$

We use the solution $x_2(\ej)$ given by~(\ref{eq:x2}) and need to carefully address its limiting behavior in $\lambda_2$, since the dominating terms cancel.
The first-order Taylor approximation of $\arctan(x)$ is $\arctan(x)=x + \mathcal{O}(x^2), \forall |x|\leq 1$. Thus, 
\begin{equation}\label{eq:arctanapprox}
\begin{split}
&\phi_1(\ej) = \arctan\bigg(\frac{\sqr{-\Xi(\ej)}}{q(\ej)}\bigg) \\
&= \frac{3\sqr{3}}{4\sx\lambda_2}\sqr{1+4\sxx{2}\lambda_1^2}  + \mathcal{O}\left(\frac{\lambda_1^3}{\sqr{\lambda_2^{3}}}\right),
\end{split}
\end{equation}
where the approximation becomes an equality in the limit as $\lambda_2 \to \infty$ since this implies that $\phi_1(\ej)\to 0$.
Similarly, for all $x$,
\begin{align*}\label{eq:cosapprox}
\cos(x/3)+\sqr{3}\sin(x/3) &= 1 + \frac{\sqr{3}}{3}x + \mathcal{O}(x^2) \\
(\cos(x/3)+\sqr{3}\sin(x/3))^2 &= 1 + \frac{2\sqr{3}}{3}x + \mathcal{O}(x^2).
\end{align*}
Let $\alpha(\ej)=\cos(\phi_1(\ej)/3) + \sqr{3}\sin(\phi_1(\ej)/3)$. Then, we can write
\begin{align}
x_2(\ej) & = -\sqr{|p(\ej)|}\alpha(\ej) -\frac{a_2(\ej)}{3} \\ \label{eq:x2sol1}
& = \frac{|p(\ej)|\alpha(\ej)^2 - \frac{a_2(\ej)^2}{9}}{-\sqr{|p(\ej)|}\alpha(\ej) + \frac{a_2(\ej)}{3}}.
\end{align}
From~(\ref{eq:cosapprox}) and using~(\ref{eq:arctanapprox}), it follows that 
\begin{equation}
\alpha(\ej)^2 
= 1 + \frac{3}{2\sx\lambda_2}\sqr{1+4\sxx{2}\lambda_1^2} + \mathcal{O}\left( \frac{\lambda_1^3}{\sqr{\lambda_2^{3}}}\right).
\end{equation}
With this, we can write the numerator of~(\ref{eq:x2sol1}) as 
\begin{align} \nonumber
&|p(\ej)|\alpha(\ej)^2 - \frac{a_2(\ej)^2}{9}  =
-\frac{\sx\lambda_2}{6\lambda_1^2} \\ \label{eq:x2nom}
&\times\big( 
2\sx\lambda_1
+1 -\sqr{1+4\sxx{2}\lambda_1^2} \big) + \mathcal{O}\big(\frac{\lambda_1}{\sqr{\lambda_2}}\big).
\end{align}
On the other hand, since $\lim_{\lambda_2\to\infty}\alpha(\ej) = 1, \forall \omega$, the denominator of~(\ref{eq:x2sol1}) can be written as (for large $\lambda_2$)
\begin{equation}\label{eq:x2den}
-\sqr{|p(\ej)|}\alpha(\ej) + \frac{a_2(\ej)}{3} \approx - \frac{2\sx\lambda_2}{3\lambda_1}.
\end{equation}
Substituting~(\ref{eq:x2nom}) and~(\ref{eq:x2den}) into~(\ref{eq:x2sol1}) yields
\begin{equation}\label{eq:dsconstant1}
\begin{split}
\Theta_-(e^{j\omega})  &= \frac{1}{4\lambda_1}\big( 
2\sx\lambda_1 + 1 -\sqr{1+4\sxx{2}\lambda_1^2} \big)   \\
& \quad + \mathcal{O}\left(\frac{\lambda_1^2}{\sqr{\lambda_2^3}}\right),
\end{split}
\end{equation}
so that 
\begin{equation}
\begin{split}
&\lim_{\lambda_2\to\infty} \Theta_-(e^{j\omega}) =
\lim_{\lambda_2\to\infty}x_2(\ej) \\
& \quad = \frac{1}{4\lambda_1}\big( 
2\sx\lambda_1 + 1 -\sqr{1+4\sxx{2}\lambda_1^2} \big).
\end{split}
\end{equation}

\subsection{Case $\lambda_1,\lambda_2\gg 1$}
We note that when assuming $\lambda_1/\sqrr{\lambda_2}\to 0$, then the results for finite $\lambda_1$ in Section~\ref{sec:case1} remain valid. 
We rewrite~(\ref{eq:dsconstant1}) as
\begin{align}\notag
\Theta_-(e^{j\omega}) &=
 \frac{1}{4\lambda_1}\,\frac{4\sx\lambda_1}{\sqr{1+4\sxx{2}\lambda_1^2}
   +2\sx\lambda_1+1} \\ \nonumber
& \quad +  \mathcal{O}\left(\frac{\lambda_1^2}{\sqr{\lambda_2^3}}\right) \\ \label{eq:theta-order}
&=\frac{1}{4\lambda_1}c_{\lambda_1}(\ej) + \mathcal{O}\left(\frac{\lambda_1^2}{\sqr{\lambda_2^3}}\right),
\end{align}
where $c_{\lambda_1}(\ej)=\mathcal{O}(1)$ and $\lim_{\lambda_1\to\infty} c_{\lambda_1}(\ej) = 1, \forall \omega$.
Inserting this into~(\ref{eq:theta1}) yields
\begin{equation}\label{eq:theta-order1}
\begin{split}
\Theta_+(e^{j\omega}) &= \frac{1}{4(\lambda_1+\lambda_2) +
  \mathcal{O}\left(\frac{\lambda_1^2}{\sqr{\lambda_2^3}}\right)}  \\
&
\quad + \mathcal{O}\left(\frac{\lambda_1^2}{\lambda_1\sqrt{\lambda_2^3} + \lambda_2\sqrt{\lambda_2^3}} +\frac{1}{\lambda_1^2 + \lambda_1\lambda_2}\right). 
\end{split}
\end{equation}
Finally, it follows from~(\ref{eq:theta-order}) and~(\ref{eq:theta-order1}) that
\begin{equation}\label{eq:hr1}
\mathop{\lim_{\lambda_1,\lambda_2\to\infty}}_{\lambda_1/\sqrr{\lambda_2}\to 0}  \lambda_1 \Theta_-(e^{j\omega})  = \frac{1}{4},
\end{equation}
and
\begin{equation}\label{eq:hr2}
\mathop{\lim_{\lambda_1,\lambda_2\to\infty}}_{\lambda_1/\sqrr{\lambda_2}\to 0} (\lambda_1+\lambda_2)  \Theta_+(e^{j\omega}) = \frac{1}{4}.
\end{equation}

\hfill \IEEEQED

\section*{Acknowledgment}
The authors would like to thank the associate editor and the referees for their many valuable suggestions that
helped improve the manuscript. The authors would also like to thank
Adam Mashiach for his comments on an earlier draft version of this manuscript.

\ifCLASSOPTIONcaptionsoff
  \newpage
\fi



\bibliographystyle{IEEEtran}
\bibliography{strings,misc,mdc,ostergaard}

\begin{thebibliography}{10}
\providecommand{\url}[1]{#1}
\csname url@samestyle\endcsname
\providecommand{\newblock}{\relax}
\providecommand{\bibinfo}[2]{#2}
\providecommand{\BIBentrySTDinterwordspacing}{\spaceskip=0pt\relax}
\providecommand{\BIBentryALTinterwordstretchfactor}{4}
\providecommand{\BIBentryALTinterwordspacing}{\spaceskip=\fontdimen2\font plus
\BIBentryALTinterwordstretchfactor\fontdimen3\font minus
  \fontdimen4\font\relax}
\providecommand{\BIBforeignlanguage}[2]{{%
\expandafter\ifx\csname l@#1\endcsname\relax
\typeout{** WARNING: IEEEtran.bst: No hyphenation pattern has been}%
\typeout{** loaded for the language `#1'. Using the pattern for}%
\typeout{** the default language instead.}%
\else
\language=\csname l@#1\endcsname
\fi
#2}}
\providecommand{\BIBdecl}{\relax}
\BIBdecl

\bibitem{elgamal:1982}
A.~A.~E. Gamal and T.~M. Cover, ``Achievable rates for multiple descriptions,''
  \emph{IEEE Trans. Inf. Theory}, vol. IT-28, no.~6, pp. 851 -- 857, November
  1982.

\bibitem{ozarow:1980}
L.~Ozarow, ``On a source-coding problem with two channels and three
  receivers,'' \emph{Bell System Technical Journal}, vol.~59, pp. 1909 -- 1921,
  December 1980.

\bibitem{Dragotti:2002}
P.~L. Dragotti, S.~D. Servetto, and M.~Vetterli, ``Optimal filter banks for
  multiple description coding: Analysis and synthesis,'' \emph{IEEE Trans. Inf.
  Theory}, vol.~48, no.~7, pp. 2036 -- 2052, July 2002.

\bibitem{chen:2009}
J.~Chen, C.~Tian, and S.~Diggavi, ``Multiple description coding for stationary
  {G}aussian sources,'' \emph{IEEE Trans. Inf. Theory}, vol.~55, no.~6, pp.
  2868--2881, June 2009.

\bibitem{kolmogorov:1956}
A.~Kolmogorov, ``On the {S}hannon theory of information transmission in the
  case of continuous signals,'' \emph{{IRE} Trans.\ Inf.\ Theory.}, vol.
  {IT}-2, pp. 102--108, 1956.

\bibitem{ostergaard:2009}
J.~{\O}stergaard and R.~Zamir, ``Multiple-description coding by dithered
  delta-sigma quantization,'' \emph{IEEE Trans. Inf. Theory}, vol.~55, no.~10,
  pp. 4661--4675, October 2009.

\bibitem{zamir:2000}
R.~Zamir, ``Shannon type bounds for multiple descriptions of a stationary
  source,'' \emph{Journal of Combinatorics, Information and System Sciences},
  pp. 1 -- 15, December 2000.

\bibitem{zamir:2008}
R.~Zamir, Y.~Kochman, and U.~Erez, ``Achieving the {G}aussian rate-distortion
  function by prediction,'' \emph{IEEE Trans. Inf. Theory}, vol.~54, no.~7, pp.
  3354 -- 3364, july 2008.

\bibitem{ingle:1995}
A.~Ingle and V.~A. Vaishampayan, ``{DPCM} system design for diversity systems
  with applications to packetized speech,'' \emph{IEEE Trans. Speech and Audio
  Proc.}, vol.~3, no.~1, January 1995.

\bibitem{regunthan:2000}
S.~L. Regunathan and K.~Rose, ``Efficient prediction in multiple description
  video coding,'' in \emph{Proc. Int. Conf. on Image Proc.}, vol.~1, 2000, pp.
  1020 -- 1023.

\bibitem{vaishampayan:1999}
V.~A. Vaishampayan and S.~John, ``Balanced interframe multiple description
  video compression,'' in \emph{Proc. Int. Conf. on Image Proc.}, vol.~3, 1999,
  pp. 812 -- 816.

\bibitem{nathan:2001}
R.~Nathan and R.~Zamir, ``Multiple description video coding with un-quantized
  prediction loop,'' in \emph{Proc. Int. Conf. on Image Proc.}, vol.~1, 2001,
  pp. 982 -- 985.

\bibitem{van_trees:1968}
H.~L. {Van Trees}, \emph{Detection, estimation, and modulation theory}.\hskip
  1em plus 0.5em minus 0.4em\relax Wiley, NY, 1968.

\bibitem{makhoul:1975}
J.~Makhoul, ``Linear prediction: A tutorial review,'' \emph{Proceedings of the
  {IEEE}}, vol.~63, no.~4, pp. 561 -- 580, April 1975.

\bibitem{zamir:1999}
R.~Zamir, ``Gaussian codes and shannon bounds for multiple descriptions,''
  \emph{IEEE Trans. Inf. Theory}, vol.~45, no.~7, pp. 2629 -- 2636, November
  1999.

\bibitem{gallager:1968}
R.~G. Gallager, \emph{Information theory and reliable communication}.\hskip 1em
  plus 0.5em minus 0.4em\relax Wiley, NY, 1968.

\bibitem{zamir:1996}
R.~Zamir and M.~Feder, ``Information rates of pre/post filtered dithered
  quantizers,'' \emph{IEEE Trans. Info. Theory}, vol.~42, no.~5, pp. 1340 --
  1353, September 1996.

\bibitem{weinstock:1974}
R.~Weinstock, \emph{Variational calculus}.\hskip 1em plus 0.5em minus
  0.4em\relax Dover Publications, 1974.

\bibitem{berger:1971}
T.~Berger, \emph{Rate {D}istortion {T}heory: {A} {M}athematical {B}asis for
  {D}ata {C}ompression}.\hskip 1em plus 0.5em minus 0.4em\relax Englewood
  Cliffs, NJ: Prentice-Hall, 1971.

\bibitem{CoverBook}
T.~M. Cover and J.~A. Thomas, \emph{Elements of Information Theory}.\hskip 1em
  plus 0.5em minus 0.4em\relax New York: Wiley, 1991.

\bibitem{boyd:2004}
S.~Boyd and L.~Vandenberghe, \emph{Convex Optimization}.\hskip 1em plus 0.5em
  minus 0.4em\relax Cambridge University Press, 2004.

\bibitem{grenander:1984}
U.~Grenander and G.~Szeg\"o, \emph{Toeplitz Forms and Their
  Applications}.\hskip 1em plus 0.5em minus 0.4em\relax New York: Chelsea,
  1984.

\bibitem{jayant:1984}
N.~S. Jayant and P.~Noll, \emph{Digital coding of waveforms}.\hskip 1em plus
  0.5em minus 0.4em\relax Englewoods {C}liffs, NJ: Prentice-{H}all, 1984.

\bibitem{zamir:1996a}
R.~Zamir and M.~Feder, ``On lattice quantization noise,'' \emph{IEEE Trans.
  Info. Theory}, vol.~42, no.~4, pp. 1152 -- 1159, July 1996.

\bibitem{erez:2004}
U.~Erez and R.~Zamir, ``Achieving 0.5 log(1+snr) over the additive white
  {G}aussian noise channel with lattice encoding and decoding,'' \emph{IEEE
  Trans. Inf. Theory}, vol.~50, pp. 2293 -- 2314, October 2004.

\bibitem{forney:2003}
G.~D. Forney, ``On the role of {MMSE} estimation in approaching the
  information-theoretic limits of linear {G}aussian channels: {S}hannon meets
  {W}iener,'' in \emph{Proc. 2003 {A}llerton Conf}, October 2003.

\bibitem{abramovitz:1973}
M.~Abramovitz and I.~A. Stegun, Eds., \emph{Handbook of mathematical
  functions}, 9th~ed.\hskip 1em plus 0.5em minus 0.4em\relax Dover
  Publications, 1973.

\end{thebibliography}

\begin{IEEEbiographynophoto}{Jan {\O}stergaard}
(S'98-M'99-SM'11) received the M.Sc.E.E.\ from Aalborg University,
Aalborg, Denmark, in 1999 and the PhD degree (cum laude) from Delft
University of Technology, Delft, The Netherlands, in 2007. From 1999
to 2002, he worked as an R\&D Engineer at ETI A/S, Aalborg, Denmark,
and from 2002 to 2003, he worked as an R\&D Engineer at ETI Inc.,
Virginia, United States. Between September 2007 and June 2008, he
worked as a post-doctoral researcher at The University of Newcastle,
NSW, Australia. From June 2008 to March 2011, he worked as a
post-doctoral researcher/Assistant Professor at Aalborg
University. Since 2011 he has been an Associate Professor at Aalborg
University. 
%

Jan {\O}stergaard has been a visiting researcher at Tel Aviv University, Tel Aviv, Israel, and at Universidad T\'ecnica Federico Santa Mar\'ia, Valpara\'iso, Chile. He has received a Danish Independent Research Council's Young Researcher's Award, a best PhD thesis award by the European Association for Signal Processing (EURASIP), and fellowships from the Danish Independent Research Council and the Villum Foundation's Young Investigator Programme. He is an Associate Editor of EURASIP Journal on Advances in Signal Processing.
\end{IEEEbiographynophoto}

\vfill
\newpage 

\begin{IEEEbiographynophoto}{Yuval Kochman}
(S'06-M'09) received his B.Sc. (cum laude), M.Sc. (cum laude) and Ph.D. degrees from Tel Aviv University in 1993, 2003 and 2010, respectively, all in electrical engineering. During 2009-2011, he was a Postdoctoral Associate at the Signals, Informtion and Algorithms Laboratory at the Massachusetts Institute of Technology (MIT), Cambridge. Since 2012, he has been with the School of Computer Science and Engineering at the Hebrew University of Jerusalem. Outside academia, he has worked in the areas of radar and digital communications. His research interests include information theory, communications and signal processing.
\end{IEEEbiographynophoto}

\begin{IEEEbiographynophoto}{Ram Zamir}
was born in Ramat-Gan, Israel in 1961. He received the B.Sc.,
M.Sc. (summa cum laude) and D.Sc. (with distinction)
degrees from Tel-Aviv University, Israel, in 1983, 1991, and 1994,
respectively, all in electrical engineering. In the years 1994 - 1996 he
spent a post-doctoral period at Cornell University,
Ithaca, NY, and at the University of California, Santa Barbara.
In 2002 he spent a Sabbatical year at MIT, and in 2008 and 2009 short
Sabbaticals at ETH and MIT. Since 1996 he has been with the department of
Elect. Eng. - Systems at Tel Aviv University.

Ram Zamir has been consulting in the areas of radar and
communications (DSL and WiFi), where he was invloved with companies like Orckit and Actelis.
During the period 2005-2014 he was the Chief Scientist of Celeno Communications.
He has been teaching information theory, data compression, random processes,
communications systems and communications circuits at Tel Aviv University.
He is an IEEE fellow since 2010.  He served as an Associate Editor for Source
Coding in the IEEE transactions on Information Theory (2001-2003),
headed the Information Theory Chapter of the Israeli IEEE
society (2000-2005), and was a member of the BOG of the society
(2013-2015).  His research interests include information theory (in particular: lattice
codes for multi-terminal problems), source coding, communications and
statistical signal processing.  His book "Lattice coding for signals
and networks" was published in 2014.
\end{IEEEbiographynophoto}

\vfill

\end{document}